\documentclass[a4paper, 11pt]{article}
\pdfoutput=1

\usepackage{jheppub}

\ifx\pdfoutput\undefined
\usepackage[dvips,bookmarks=false]{hyperref}    %This is for arXiv.org
\else
\usepackage{hyperref}   %This is for pdftex
\hypersetup{colorlinks,bookmarksopen,bookmarksnumbered,citecolor=blue,linkcolor=blue,pdfstartview=FitH,urlcolor=blue}
\fi

\newcommand{\mycaption}{\caption}

\newcommand{\parenbar}[1]{{\overset{%
            \protect\raisebox{-0.15em}{\protect\scalebox{.4}{\textbf{(}}}%
            \protect\raisebox{-0.3em}{{\hspace{.03em}--\hspace{.05em}}}%
            \protect\raisebox{-0.15em}{\protect\scalebox{.4}{\textbf{)}}}} {#1}}}
\newcommand{\pbn}{\parenbar{\nu}}
\newcommand{\iso}[2]{{\ensuremath{\mbox{}^{#2}}\ensuremath{\text{#1}}}}

\newcommand{\Dmq}{\ensuremath{\Delta m^2}}
\newcommand{\eVq}{\ensuremath{\text{eV}^2}}
\newcommand{\diag}{\mathop{\mathrm{diag}}}
\newcommand{\sumSte}{\sum_{\sigma=\text{st}}}
\newcommand{\sumAll}{\sum_{i=\text{all}}}
\renewcommand{\Re}{\mathop{\mathrm{Re}}}
\renewcommand{\Im}{\mathop{\mathrm{Im}}}
\newcommand{\Evol}{\mathop{\mathrm{Evol}}}

\newcommand{\fitnumber}[1]{{#1}}

\makeatletter
\DeclareRobustCommand\recite[1]{\begingroup\@fileswfalse\cite{#1}\endgroup}
\makeatother

\title{Sterile neutrino oscillations: the global picture}

\author[a]{Joachim Kopp,}
\affiliation[a]{Max-Planck-Institut f\"ur Kernphysik,
  Saupfercheckweg 1, 69117 Heidelberg, Germany}
\emailAdd{jkopp@mpi-hd.mpg.de}

\author[b,c]{Pedro A.N.\ Machado,}
\affiliation[b]{Instituto de F\'{i}sica, Universidade de S\~{a}o Paulo,\\
    C.P.~66.318, 05315-970 S\~{a}o Paulo, Brazil}
\affiliation[c]{Institut de Physique Th\'eorique,\\
    CEA-Saclay, 91191 Gif-sur-Yvette, France}
\emailAdd{accioly@fma.if.usp.br}

\author[d]{Michele Maltoni,}
\affiliation[d]{Instituto de F\'isica Te\'orica UAM/CSIC,\\
    Calle de Nicol\'as Cabrera 13-15, E-28049 Madrid, Spain}
\emailAdd{michele.maltoni@csic.es}

\author[a]{Thomas Schwetz}
\emailAdd{schwetz@mpi-hd.mpg.de}

\abstract{Neutrino oscillations involving eV-scale neutrino mass
  states are investigated in the context of global neutrino
  oscillation data including short and long-baseline accelerator,
  reactor, and radioactive source experiments, as well as atmospheric
  and solar neutrinos. We consider sterile neutrino mass schemes
  involving one or two mass-squared differences at the \eVq\ scale
  denoted by 3+1, 3+2, and 1+3+1. We discuss the hints for eV-scale
  neutrinos from $\pbn_e$ disappearance (reactor and Gallium
  anomalies) and $\pbn_\mu\to\pbn_e$ appearance (LSND and MiniBooNE)
  searches, and we present constraints on sterile neutrino mixing from
  $\pbn_\mu$ and neutral-current disappearance data.  An explanation
  of all hints in terms of oscillations suffers from severe tension
  between appearance and disappearance data. The best compatibility is
  obtained in the 1+3+1 scheme with a p-value of 0.2\% and exceedingly
  worse compatibilities in the 3+1 and 3+2 schemes.}

\preprint{IFT-UAM/CSIC-13-026}


\begin{document}

\maketitle

%==============================================================================
\section{Introduction}
\label{sec:intro}
%==============================================================================

Huge progress has been made in the study of neutrino
oscillations~\cite{Fukuda:1998mi, Ahmad:2002jz, Araki:2004mb,
  Adamson:2008zt}, and with the recent determination of the last
unknown mixing angle $\theta_{13}$~\cite{Abe:2011sj, Adamson:2011qu,
  Abe:2011fz, An:2012eh, Ahn:2012nd, Abe:2012tg} a clear first-order
picture of the three-flavor lepton mixing matrix has emerged, see
\textit{e.g.}~\cite{GonzalezGarcia:2012sz}. Besides those achievements there
are some anomalies which cannot be explained within the three-flavor
framework and which might point towards the existence of additional
neutrino flavors (so-called sterile neutrinos) with masses at the eV
scale:
\begin{itemize}
\item The LSND experiment~\cite{Aguilar:2001ty} reports evidence for
  $\bar\nu_\mu\to\bar\nu_e$ transitions with $E/L \sim 1~\eVq$,
  where $E$ and $L$ are the neutrino energy and the distance between
  source and detector, respectively.

\item This effect is also searched for by the MiniBooNE
  experiment~\cite{AguilarArevalo:2007it, AguilarArevalo:2010wv,
    MBnu2012, AguilarArevalo:2012va, Aguilar-Arevalo:2013ara}, which
  reports a yet unexplained event excess in the low-energy region of
  the electron neutrino and anti-neutrino event spectra. No
  significant excess is found at higher neutrino energies.
  Interpreting the data in terms of oscillations, parameter values
  consistent with the ones from LSND are obtained.

\item Radioactive source experiments at the Gallium solar neutrino
  experiments SAGE and GALLEX have obtained an event rate which is
  somewhat lower than expected. This effect can be explained by the
  hypothesis of $\nu_e$ disappearance due to oscillations with $\Dmq
  \gtrsim 1~\eVq$~\cite{Acero:2007su, Giunti:2010zu} (``Gallium
  anomaly'').

\item A recent re-evaluation of the neutrino flux emitted by nuclear
  reactors~\cite{Mueller:2011nm, Huber:2011wv} has led to somewhat
  increased fluxes compared to previous
  calculations~\cite{Schreckenbach:1985ep, Hahn:1989zr,
    VonFeilitzsch:1982jw, Vogel:1980bk}. Based on the new flux
  calculation, the results of previous short-baseline ($L \lesssim
  100$~m) reactor experiments are in tension with the prediction, a
  result which can be explained by assuming $\bar\nu_e$ disappearance
  due to oscillations with $\Dmq \sim 1~\eVq$~\cite{Mention:2011rk}
  (``reactor anomaly'').
\end{itemize}
Sterile neutrino oscillation schemes have been considered for a long
time, see \textit{e.g.}~\cite{GomezCadenas:1995sj, Goswami:1995yq,
  Bilenky:1996rw,Okada:1996kw} for early references on four-neutrino
scenarios. Effects of two sterile neutrinos at the eV scale have been
considered first in~\cite{Peres:2000ic, Sorel:2003hf}, oscillations
with three sterile neutrinos have been investigated
in~\cite{Maltoni:2007zf, Conrad:2012qt}.

Thus, while the phenomenology of sterile neutrino models is well
known, it has also been known for a long time that the LSND and
MiniBooNE $\parenbar\nu_e$ appearance signals are in tension with
bounds from disappearance experiments~\cite{Maltoni:2002xd,
  Strumia:2002fw, Cirelli:2004cz}, challenging an interpretation in
terms of sterile neutrino oscillations. This problem remains severe,
and in the following we will give a detailed discussion of the status
of the $\parenbar{\nu}_\mu \to \parenbar{\nu}_e$ appearance hints from
LSND and MiniBooNE in the light of recent global data. The situation
is better for the hints for $\parenbar{\nu}_e$ disappearance from the
reactor and Gallium anomalies, which are not in direct conflict with
any other data. This somewhat ambiguous situation asks for an
experimental answer, and indeed several projects are under preparation
or under investigation, ranging from experiments with radioactive
sources, short-baseline reactor experiments, to new accelerator
facilities. A recent review on light sterile neutrinos including an
overview on possible experimental tests can be found
in~\cite{Abazajian:2012ys}.

In this paper we provide an extensive analysis of the present
situation of sterile neutrino scenarios. We discuss the possibility to
explain the tentative positive signals from LSND and MiniBooNE, as
well as the reactor and Gallium anomalies in terms of sterile neutrino
oscillations in view of the global data.  New ingredients with respect
to our previous analysis~\cite{Kopp:2011qd} are the following.
\begin{itemize}
\item We use latest data from the MiniBooNE $\parenbar{\nu}_\mu \to
  \parenbar{\nu}_e$ appearance searches~\cite{MBnu2012,
    AguilarArevalo:2012va, Aguilar-Arevalo:2013ara}.  Our MiniBooNE
  appearance analysis is now based on Monte Carlo events provided by
  the collaboration taking into account realistic event
  reconstruction, correlation matrices, as well as oscillations of
  various background components in a consistent way.

\item We include the constraints on the appearance probability from
  E776~\cite{Borodovsky:1992pn} and ICARUS~\cite{Antonello:2012pq}.

\item We include the Gallium anomaly in our fit.

\item We take into account constraints from solar neutrinos, the
  KamLAND reactor experiment, and LSND and KARMEN measurements of the
  reaction $\nu_e + \iso{C}{12} \to e^- + \iso{N}{12}$.

\item The treatment of the reactor anomaly is improved
and updated by taking into account small changes in the predicted
anti-neutrino fluxes as well as an improved consideration of
systematic errors and their correlations.

\item We take into account charged-current (CC) and neutral-current
  (NC) data from the MINOS long-baseline
  experiment~\cite{Adamson:2010wi, Adamson:2011ku}.

\item We include data on $\nu_\mu$ disappearance from
  MiniBooNE~\cite{AguilarArevalo:2009yj} as well as $\bar\nu_\mu$
  disappearance from a joint MiniBooNE/SciBooNE
  analysis~\cite{Cheng:2012yy}.

\item In our analysis of atmospheric neutrino data, we improve our
  formalism to fully take into account the mixing of $\nu_e$ with
  other active or sterile neutrino states.
\end{itemize}
All the data used in this work are summarized in
Tab.~\ref{tab:data-summary}. For other recent sterile neutrino global
fits see~\cite{Conrad:2012qt, Giunti:2011hn, Archidiacono:2013xxa}. We
are restricting our analysis to neutrino oscillation data;
implications for kinematic neutrino mass measurements and
neutrino-less double beta-decay data have been discussed recently
in~\cite{Li:2011ss, Barry:2011wb, Giunti:2011cp}.

\begin{table}[t]
  \centering\footnotesize
  \begin{tabular}{lrcr}
    \hline\hline
    Experiment & dof & channel & comments \\
    \hline
    Short-baseline reactors  &  76 & $\bar\nu_e \to \bar\nu_e$ & SBL \\
    Long-baseline reactors   &  39 & $\bar\nu_e \to \bar\nu_e$ & LBL\\
    KamLAND                  &  17 & $\bar\nu_e \to \bar\nu_e$ & \\
    Gallium                  &   4 & $\nu_e \to \nu_e$ & SBL\\
    Solar neutrinos          & 261 & \multicolumn{2}{c}{$\nu_e \to \nu_e$ + NC data} \\
    LSND/KARMEN \iso{C}{12}  & 32 & $\nu_e \to \nu_e$ & SBL\\
    \hline
    CDHS                     &  15 & $\nu_\mu \to \nu_\mu$ & SBL \\
    MiniBooNE $\nu$          &  15 & $\nu_\mu \to \nu_\mu$ & SBL \\
    MiniBooNE $\bar\nu$      & 42 & $\bar{\nu}_\mu \to \bar{\nu}_\mu$ & SBL \\
    MINOS CC                 &  20 & $\nu_\mu \to \nu_\mu$ & LBL  \\
    MINOS NC                 &  20 & $\nu_\mu \to \nu_s$ & LBL  \\
    Atmospheric neutrinos    & 80 & \multicolumn{2}{c}{$\parenbar{\nu}_\mu \to \parenbar{\nu}_\mu$ + NC matter effect} \\
    \hline
    LSND                     &  11 & $\bar\nu_\mu\to\bar\nu_e$ & SBL \\
    KARMEN                   &  9  & $\bar\nu_\mu\to\bar\nu_e$ &SBL \\
    NOMAD                    &  1  & $\nu_\mu\to\nu_e$ &SBL \\
    MiniBooNE $\nu$          &  11 & $\nu_\mu \to {\nu}_e$ &SBL \\
    MiniBooNE $\bar\nu$      &  11 & $\bar\nu_\mu \to \bar\nu_e$ &SBL \\
    E776                     &  24 & $\parenbar{\nu}_\mu \to \parenbar{\nu}_e$ &SBL \\
    ICARUS                   &   1 & $\nu_\mu\to\nu_e$ & LBL \\
    \hline
    total                    & 689 & &\\
    \hline\hline
  \end{tabular}
  \mycaption{Summary of the data used in this work divided into
    $\pbn_e$, $\pbn_\mu$ disappearance, and appearance data. The
    column ``dof'' gives the number of data points used in our
    analysis minus the number of free nuisance parameters for each
    experiment.\label{tab:data-summary}}
\end{table}

Sterile neutrinos at the eV scale also have implications for
cosmology. If thermalized in the early Universe they contribute to the
number of relativistic degrees of freedom (effective number of
neutrino species $N_\text{eff}$). A review with many references can be
found in~\cite{Abazajian:2012ys}.  Indeed there might be some hints
from cosmology for additional relativistic degrees of freedoms
($N_\text{eff}$ bigger than 3), coming mainly from CMB data,
\textit{e.g.}~\cite{Hamann:2010bk, Giusarma:2011ex,
  GonzalezGarcia:2010un, Archidiacono:2012ri, NeffLunardini:2013,
  Archidiacono:2013xxa}.  Recently precise CMB data from the PLANCK
satellite have been released~\cite{Ade:2013zuv}. Depending on which
additional cosmological data are used, $N_\text{eff}$ values ranging
from $3.30^{+0.54}_{-0.51}$ to $3.62^{+0.50}_{-0.48}$ (uncertainties
at 95\%~CL) are obtained~\cite{Ade:2013zuv}. Constraints from Big Bang
Nucleosynthesis on $N_\text{eff}$ have been considered recently
in~\cite{Mangano:2011ar}.
Apart from their contribution to $N_\text{eff}$, thermalized eV-scale
neutrinos would also give a large contribution to the sum of neutrino
masses, which is constrained to be below around 0.5~eV. The exact
constraint depends on which cosmological data sets are used, but the
most important observables are those related to galaxy
clustering~\cite{Hamann:2010bk, Giusarma:2011ex,
  GonzalezGarcia:2010un, Archidiacono:2012ri}. In the standard
$\Lambda$CDM cosmology framework the bound on the sum of neutrino
masses is in tension with the masses required to explain the
aforementioned terrestrial hints~\cite{Archidiacono:2012ri}. The
question to what extent such sterile neutrino scenarios are disfavored
by cosmology and how far one would need to deviate from the
$\Lambda$CDM model in order to accommodate them remains under
discussion~\cite{Hamann:2011ge, Joudaki:2012uk,
  Archidiacono:2013xxa}. We will not include any information from
cosmology explicitly in our numerical analysis. However, we will keep
in mind that neutrino masses in excess of few eV may become more and
more difficult to reconcile with cosmological observations.

The outline of the paper is as follows. In
Sec.~\ref{sec:parameterisation} we introduce the formalism of sterile
neutrino oscillations and fix the parametrization of the mixing
matrix.  We then consider $\parenbar\nu_e$ disappearance data in
Sec.~\ref{sec:nue-dis}, discussing the reactor and Gallium
anomalies. Constraints from $\parenbar\nu_\mu$ disappearance as well
as neutral-current data are discussed in Sec.~\ref{sec:numu-dis}, and
global $\parenbar\nu_\mu \to \parenbar\nu_e$ appearance data including
the LSND and MiniBooNE signals in Sec.~\ref{sec:nue-app}.  The global
fit of all these data combined is presented in Sec.~\ref{sec:combi}
for scenarios with one or two sterile neutrinos. We summarize our
results and conclude in Sec.~\ref{sec:conclusions}. Supplementary
material is provided in the appendices including a discussion of
complex phases in sterile neutrino oscillations, oscillation
probabilities for solar and atmospheric neutrinos, as well as
technical details of our experiment simulations.

% =============================================================================
\section{Oscillation parameters in the presence of sterile neutrinos}
\label{sec:parameterisation}
% =============================================================================

In this work we consider the presence of $s = 1$ or 2 additional
neutrino states with masses in the $\lesssim$~few~eV range. When
moving from 1 to 2 sterile neutrinos the qualitative new feature is
the possibility of CP violation already at
short-baseline~\cite{Karagiorgi:2006jf,
  Maltoni:2007zf}.\footnote{Adding more than two sterile neutrinos
  does not lead to any qualitatively new physical effects and as shown
  in~\cite{Maltoni:2007zf} the fit does not improve
  significantly. Therefore, we restrict the present analysis to $s \le
  2$ sterile neutrinos.}  The neutrino mass eigenstates
$\nu_1,\ \dots,\ \nu_{3+s}$ are labeled such that $\nu_1$, $\nu_2$,
$\nu_3$ contribute mostly to the active flavor eigenstates and provide
the mass squared differences required for ``standard'' three-flavor
oscillations, $\Dmq_{21} \approx 7.5\times 10^{-5}~\eVq$ and
$|\Dmq_{31}| \approx 2.4\times 10^{-3}~\eVq$, where $\Dmq_{ij} \equiv
m_i^2 - m_j^2$.  The mass states $\nu_4$, $\nu_5$ are mostly sterile
and provide mass-squared differences in the range $0.1~\eVq \lesssim
|\Dmq_{41}|, |\Dmq_{51}| \lesssim 10~\eVq$. In the case of only one
sterile neutrino, denoted by ``3+1'' in the following, we always
assume $\Dmq_{41} > 0$, but the oscillation phenomenology for $\Delta
m_{41}^2 < 0$ would be the same. For two sterile neutrinos, we
distinguish between a mass spectrum where $\Dmq_{41}$ and $\Dmq_{51}$
are both positive (``3+2'') and where one of them is negative
(``1+3+1'').  The phenomenology is slightly different in the two
cases~\cite{Goswami:2007kv}.
We assume that the $s$ linear combinations of mass states which are
orthogonal to the three flavor states participating in weak
interactions are true singlets and have no interaction with Standard
Model particles. Oscillation physics is then described by a
rectangular mixing matrix $U_{\alpha i}$ with $\alpha = e,\mu,\tau$
and $i=1,...,3+s$, and $\sum_i U_{\alpha i}^* U_{\beta i} =
\delta_{\alpha\beta}$.\footnote{In this work we consider so-called
  phenomenological sterile neutrino models, where the $3+s$ neutrino
  mass eigenvalues and the mixing parameters $U_{\alpha i}$ are
  considered to be completely independent. In particular we do not
  assume a seesaw scenario, where the Dirac and Majorana mass matrices
  of the sterile neutrinos are the only source of neutrino mass and
  mixing. For such ``minimal'' sterile neutrino models see
  \textit{e.g.}~\cite{Blennow:2011vn, Fan:2012ca,Donini:2012tt}.}

We give here expressions for the oscillation probabilities in vacuum,
focusing on the 3+2 case. It is trivial to recover the 3+1 formulas
from them by simply dropping all terms involving the index
``5''. Formulas for the 1+3+1 scenario are obtained by taking either
$\Dmq_{51}$ or $\Dmq_{41}$ negative. Oscillation probabilities
relevant for solar and atmospheric neutrinos are given in
appendices~\ref{app:solar} and~\ref{app:atm}, respectively.

First we consider the so-called ``short-baseline'' (SBL) limit, where
the relevant range of neutrino energies and baselines is such that
effects of $\Dmq_{21}$ and $\Dmq_{31}$ can be neglected. Then,
oscillation probabilities depend only on $\Dmq_{i1}$ and $U_{\alpha
  i}$ with $i\geq 4$.  We obtain for the appearance probability
\begin{multline} \label{eq:SBLapp}
  P_{\nu_\alpha\to\nu_\beta}^\text{SBL,3+2} =
  4 \, |U_{\alpha 4}|^2 |U_{\beta 4}|^2 \, \sin^2 \phi_{41} +
  4 \, |U_{\alpha 5}|^2 |U_{\beta 5}|^2 \, \sin^2 \phi_{51}
  \\
  + 8 \,|U_{\alpha 4}U_{\beta 4}U_{\alpha 5}U_{\beta 5}| \,
  \sin\phi_{41}\sin\phi_{51}\cos(\phi_{54} - \gamma_{\alpha\beta}) \,,
\end{multline}
with the definitions
\begin{equation} \label{eq:5nu-def}
  \phi_{ij} \equiv \frac{\Dmq_{ij}L}{4E} \,,
  \qquad \gamma_{\alpha\beta} \equiv
  \arg\left(I_{\alpha\beta 54} \right)
  \,,\qquad
  I_{\alpha\beta ij} \equiv U_{\alpha i}^* U_{\beta i} U_{\alpha j} U_{\beta j}^* \,.
\end{equation}
Eq.~\eqref{eq:SBLapp} holds for neutrinos; for anti-neutrinos one has
to replace $\gamma_{\alpha\beta} \to -\gamma_{\alpha\beta}$. Since
Eq.~\eqref{eq:SBLapp} is invariant under the transformation
$4\leftrightarrow 5$ and $\gamma_{\alpha\beta} \to
-\gamma_{\alpha\beta}$, we can restrict the parameter range to
$\Dmq_{54} \ge 0$, or equivalently $\Dmq_{51} \ge \Dmq_{41}$, without
loss of generality. Note also that the probability
Eq.~\eqref{eq:SBLapp} depends only on the combinations $|U_{\alpha
  4}U_{\beta 4}|$ and $|U_{\alpha 5}U_{\beta 5}|$.  The only SBL
appearance experiments we are considering are in the
$\parenbar{\nu}_\mu\to\parenbar{\nu}_e$ channel. Therefore, the total
number of independent parameters is 5 if only SBL appearance
experiments are considered.

The 3+2 survival probability, on the other hand, is given in the SBL
approximation by
\begin{equation} \label{eq:SBLdis}
  P_{\nu_\alpha\to\nu_\alpha}^\text{SBL,3+2} = 1
  - 4\left(1 - \sum_{i=4,5} |U_{\alpha i}|^2 \right)
  \sum_{i=4,5} |U_{\alpha i}|^2 \, \sin^2\phi_{i1}
  - 4\, |U_{\alpha 4}|^2|U_{\alpha 5}|^2 \, \sin^2\phi_{54} \,.
\end{equation}

In this work we include also experiments for which the SBL
approximation cannot be adopted, in particular MINOS and ICARUS. For
these experiments $\phi_{31}$ is of order one. In the following we
give the relevant oscillation probabilities in the limit of
$\phi_{41},\phi_{51},\phi_{54} \to \infty$ and $\phi_{21} \to 0$. We
call this the long-baseline (LBL) approximation. In this case we
obtain for the neutrino appearance probability ($\alpha \neq \beta$)
\begin{multline}\label{eq:LBLapp}
  P_{\nu_\alpha \to \nu_\beta}^\text{LBL,3+2} = 4|U_{\alpha 3}|^2|U_{\beta 3}|^2 \sin^2\phi_{31}
  + 2 \sum_{i=4}^5 |U_{\alpha i}|^2|U_{\beta i}|^2 + 2\Re(I_{\alpha\beta 45})
  \\
  + 4 \Re(I_{\alpha\beta 43} + I_{\alpha\beta 53})\sin^2\phi_{31}
  + 2 \Im(I_{\alpha\beta 43} + I_{\alpha\beta 53}) \sin(2\phi_{31}) \,.
\end{multline}
The corresponding expression for anti-neutrinos is obtained by the
replacement $I_{\alpha\beta ij} \to I_{\alpha\beta ij}^*$. The
survival probability in the LBL limit can be written as
\begin{equation}\label{eq:LBLdis}
  P_{\nu_\alpha\to\nu_\alpha}^\text{LBL,3+2} = \left(1 - \sum_{i=3}^5|U_{\alpha i}|^2\right)^2
  + \sum_{i=3}^5|U_{\alpha i}|^4 +
  2 \left(1 - \sum_{i=3}^5|U_{\alpha i}|^2\right) |U_{\alpha 3}|^2 \cos(2\phi_{31}) \,.
\end{equation}
Note that in the numerical analysis of MINOS data neither the SBL nor
the LBL approximations can be used because $\phi_{31}$, $\phi_{41}$
and $\phi_{51}$ can all become of order one either at the far detector
or at the near detector~\cite{Hernandez:2011rs}. Moreover, matter
effects cannot be neglected in MINOS. All of these effects are
properly included in our numerical analysis of the MINOS experiment.

Sometimes it is convenient to complete the $3\times (3+s)$ rectangular
mixing matrix by $s$ rows to an $n\times n$ unitary matrix, with
$n=3+s$. For $n=5$ we use the following parametrization for $U$:
\begin{equation}\label{eq:rotations}
  U = V_{35} O_{34}V_{25}V_{24}O_{23}O_{15}O_{14}V_{13}V_{12}
\end{equation}
where $O_{ij}$ represents a real rotation matrix by an angle
$\theta_{ij}$ in the $ij$ plane, and $V_{ij}$ represents a complex
rotation by an angle $\theta_{ij}$ and a phase $\varphi_{ij}$. The
particular ordering of the rotation matrices is an arbitrary
convention which, however, turns out to be convenient for practical
reasons.\footnote{Note that the ordering chosen in
  Eq.~\eqref{eq:rotations} is equivalent to $U = V_{35} V_{25} O_{15}
  O_{34}V_{24}O_{14}O_{23}V_{13}V_{12}$, where the standard
  three-flavor convention appears to the right (apart from an
  additional complex phase), and mixing involving the mass states
  $\nu_4$ and $\nu_5$ appear successively to the left of it.} We have
dropped the unobservable rotation matrix $V_{45}$ which just mixes
sterile states.  There is also some freedom regarding which phases are
removed by field redefinitions and which ones are kept as physical
phases.  In appendix~\ref{app:phases} we give a specific recipe for
how to remove unphysical phases in a consistent way. Throughout this
work we consider only phases which are phenomenologically relevant in
neutrino oscillations.  Under certain approximations, more phases may
become unphysical. For instance, if an angle which corresponds to a
rotation which can be chosen to be complex is zero the corresponding
phase disappears. In practical situations often one or more of the
mass-squared differences can be considered to be zero, which again
implies that some of the angles and phases will become unphysical. In
Tab.~\ref{tab:counting} we show the angle and phase counting for the
SBL and LBL approximations for the 3+2 and 3+1 cases.

\begin{table}[t]
  \centering
  \begin{tabular}{cc@{\qquad}rc@{\qquad}rc}
    \hline\hline
    & A/P & LBL approx. & (A/P) & SBL approx. & (A/P) \\
    \hline
    3+2 & 9/5 & $V_{35}V_{34}V_{25}O_{24}O_{23}O_{15}O_{14}V_{13}$ & (8/4)
              & $V_{35}O_{34}V_{25}O_{24}O_{15}O_{14}$ & (6/2) \\
    3+1 & 6/3 & $V_{34}O_{24}O_{23}O_{14}V_{13}$ & (5/2)
              & $O_{34}O_{24}O_{14}$ & (3/0) \\
    \hline\hline
  \end{tabular}
  \mycaption{Mixing angle and Phase counting for $s=2$ (3+2) and $s=1$
    (3+1) sterile neutrino schemes. The column ``A/P'' denotes the
    number of physical angles and phases, respectively. The column
    ``LBL approx.'' (``SBL approx.'')  corresponds to the
    approximation $\Dmq_{21} \to 0$ ($\Dmq_{21} \to 0$, $\Dmq_{31}\to
    0$). We give also specific examples for which angles can be chosen
    real, by denoting with $V_{ij}$ ($O_{ij}$) a complex (real)
    rotation.\label{tab:counting}}
\end{table}

In the notation of Eqs.~\eqref{eq:SBLapp}, \eqref{eq:SBLdis},
\eqref{eq:LBLapp}, \eqref{eq:LBLdis}, it is explicit that only
appearance experiments depend on complex phases in a parametrization
independent way.  However, in a particular parametrization such as
Eq.~\eqref{eq:rotations}, also the moduli $|U_{\alpha i}|$ may depend
on cosines of the phase parameters $\varphi_{ij}$, leading to some
sensitivity of disappearance experiments to the $\varphi_{ij}$ in a
CP-even fashion. Our parametrization Eq.~\eqref{eq:rotations}
guarantees that $\parenbar\nu_e$ disappearance experiments are
independent of $\varphi_{ij}$.

% =============================================================================
\section{$\nu_e$ and $\bar\nu_e$ disappearance searches}
\label{sec:nue-dis}
% =============================================================================

Disappearance experiments in the $\parenbar \nu_e$ sector probe
the moduli of the entries in the first row of the neutrino mixing matrix,
$|U_{ei}|$. In the short-baseline limit of the 3+1 scenario, the only relevant parameter
is $|U_{e4}|$. For two sterile neutrinos, also $|U_{e5}|$ is relevant.
In this section we focus on 3+1 models, and comment
only briefly on 3+2. For 3+1 oscillations in the SBL limit, the $\parenbar \nu_e$
survival probability takes an effective two flavor form
\begin{equation}
  P_{ee}^\text{SBL,3+1} =
  1 - 4 |U_{e4}|^2(1 - |U_{e4}|^2) \sin^2\frac{\Dmq_{41} L}{4E}
  = 1 - \sin^22\theta_{ee} \sin^2\frac{\Dmq_{41} L}{4E} \,,
\end{equation}
where we have defined an effective $\parenbar \nu_e$-disappearance
mixing angle by
\begin{equation}\label{eq:sinq2t_ee}
  \sin^2 2\theta_{ee} \equiv 4 |U_{e4}|^2(1 - |U_{e4}|^2)\,.
\end{equation}
This definition is parametrization independent. Using the specific
parametrization of Eq.~\eqref{eq:rotations} it turns out that
$\theta_{ee} = \theta_{14}$.

%------------------------------------------------------------------------------
\subsection{SBL reactor experiments}
%------------------------------------------------------------------------------

\begin{table}[t]
  \centering\footnotesize
  \begin{tabular}{lrccc}
    \hline\hline
    experiment & $L$ [m]& obs/pred & unc.~error [\%] & tot.~error [\%]\\
    \hline
    Bugey4~\cite{Declais:1994ma}       & 15 &0.926 &1.09& 1.37\\
    Rovno91~\cite{Kuvshinnikov:1990ry} & 18 &0.924 &2.10& 2.76\\
    Bugey3~\cite{Declais:1994su}       & 15 &0.930 &2.05& 4.40\\
    Bugey3~\cite{Declais:1994su}       & 40 &0.936 &2.06& 4.41\\
    Bugey3~\cite{Declais:1994su}       & 95 &0.861 &14.6& 1.51\\
    Gosgen~\cite{Zacek:1986cu}         & 38 &0.949 &2.38& 5.35\\
    Gosgen~\cite{Zacek:1986cu}         & 45 &0.975 &2.31& 5.32\\
    Gosgen~\cite{Zacek:1986cu}         & 65 &0.909 &4.81& 6.79\\
    ILL~\cite{Kwon:1981ua}             &  9 &0.788 &8.52& 1.16\\
    Krasnoyarsk~\cite{Vidyakin:1987ue} & 33 &0.920 &3.55& 6.00\\
    Krasnoyarsk~\cite{Vidyakin:1987ue} & 92 &0.937 &19.8& 2.03\\
    Krasnoyarsk~\cite{Vidyakin:1994ut} & 57 &0.931 &2.67& 4.32\\
    SRP~\cite{Greenwood:1996pb}        & 18 &0.936 &1.95& 2.79\\
    SRP~\cite{Greenwood:1996pb}        & 24 &1.001 &2.11& 2.90\\
    Rovno88~\cite{Afonin:1988gx}       & 18 &0.901 &4.24& 6.38\\
    Rovno88~\cite{Afonin:1988gx}       & 18 &0.932 &4.24& 6.38\\
    Rovno88~\cite{Afonin:1988gx}       & 18 &0.955 &4.95& 7.33\\
    Rovno88~\cite{Afonin:1988gx}       & 25 &0.943 &4.95& 7.33\\
    Rovno88~\cite{Afonin:1988gx}       & 18 &0.922 &4.53& 6.77\\
    \hline
    Palo Verde~\cite{Boehm:2001ik}     &  820 &\multicolumn{3}{c}{1 rate} \\
    Chooz~\cite{Apollonio:2002gd}      & 1050 &\multicolumn{3}{c}{14 bins}\\
    DoubleChooz~\cite{Abe:2012tg}      & 1050 &\multicolumn{3}{c}{18 bins} \\
    DayaBay~\cite{DBneutrino}          & &\multicolumn{3}{c}{6 rates -- 1 norm} \\
    RENO~\cite{Ahn:2012nd}             & &\multicolumn{3}{c}{2 rates -- 1 norm} \\
    \hline
    KamLAND~\cite{Gando:2010aa}        & &\multicolumn{3}{c}{17 bins} \\
    \hline\hline
  \end{tabular}
  \mycaption{Reactor data used in our analysis. The experiments in the
    upper part of the table have baselines $L < 100$~m and are
    referred to as SBL reactor experiments. For these experiments we
    list the baseline, the ratio of the observed and predicted rates
    (based on the flux predictions from~\recite{Mueller:2011nm,
      Huber:2011wv}), the uncorrelated error, and the total
    experimental error (\textit{i.e.}, the square-root of the diagonal
    entry of the correlation matrix).  Uncertainties from the neutrino
    flux prediction are not included here, but are taken into account
    in our numerical analysis. For details on the correlations and
    flux errors see appendix~\ref{app:react-errors}. In the lower part
    of the table, we list experiments with baselines of order 1~km
    (LBL reactors), and the KamLAND experiment with an average
    baseline of 180~km.  For DayaBay, RENO, and KamLAND, we do not
    give a number for the baseline here because several baselines are
    involved in each of these experiments.  The number of SBL data
    points is 19 or 76 and the total number of reactor data points is
    75 or 132, depending on whether a total rates analysis (3 data
    points) or a spectral analysis (25+25+10~bins) is used for the
    Bugey3 experiment.\label{tab:react-data}}
\end{table}

The data from reactor experiments used in our analysis are summarized
in Tab.~\ref{tab:react-data}. Our simulations make use of a dedicated
reactor code based on previous publications, see
\textit{e.g.}~\cite{Grimus:2001mn, Schwetz:2011qt}. We have updated
the code to include the latest data and improved the treatment of
uncertainties, see appendix~\ref{app:react-errors} for details. The
code used here is very similar to the one from
Ref.~\cite{GonzalezGarcia:2012sz}, extended to sterile neutrino
oscillations. The reactor experiments listed in
Tab.~\ref{tab:react-data} can be divided into short-baseline (SBL)
experiments with baselines $<100$~m, long-baseline (LBL) experiments
with $100~\text{m} < L < 2~\text{km}$, and the very long-baseline
experiment KamLAND with an average baseline of 180~km. SBL experiments
are not sensitive to standard three-flavor oscillations, but can
observe oscillatory behavior for $\Dmq_{41}$, $\Dmq_{51} \sim
1~\eVq$. On the other hand, for long-baseline experiments,
oscillations due to $(\Dmq_{31}$, $\theta_{13})$ are most relevant,
and oscillations due to \eVq-scale mass-squared differences are
averaged out and lead only to a constant flux suppression. KamLAND is
sensitive to oscillations driven by $(\Dmq_{21}$ and $\theta_{12})$,
whereas all $\theta_{1k}$ with $k \ge 3$ lead only to a constant flux
reduction.

For the SBL reactor experiments we show in Tab.~\ref{tab:react-data}
also the ratio of the observed and predicted rate, where the latter is
based on the flux calculations of~\cite{Huber:2011wv} for neutrinos
from $^{235}$U, $^{239}$Pu, $^{241}$Pu fission
and~\cite{Mueller:2011nm} for $^{238}$U fission. The ratios are taken
from~\cite{Abazajian:2012ys} (which provides and update
of~\cite{Mention:2011rk}) and are based on the Particle Data Group's
2011 value for the neutron lifetime, $\tau_n =
881.5$~s~\cite{Nakamura:2010zzi}.\footnote{This number differs from
  their 2012 value 880.1~s by less than 0.2\%~\cite{pdg}. The neutron
  lifetime enters the calculation of the detection cross section and
  therefore has a direct impact on the expected rate.  The quoted
  uncertainties of $\lesssim 0.2\%$ are small compared to the
  uncertainties on the predicted flux, see~\cite{RevModPhys.83.1173}
  for a discussion of the neutron lifetime determination.} We observe
that most of the ratios are smaller than one. In order to asses the
significance of this deviation, a careful error analysis is
necessary. In the last column of Tab.~\ref{tab:react-data}, we give
the uncorrelated errors on the rates.  They include statistical as
well as uncorrelated experimental errors. In addition to these, there
are also correlated experimental errors between various data points
which are described in detail in
appendix~\ref{app:react-errors}. Furthermore, we take into account the
uncertainty on the neutrino flux prediction following the prescription
given in~\cite{Huber:2011wv}, see also appendix~\ref{app:react-errors}
for details.

Fitting the SBL data to the predicted rates we obtain
$\chi^2/\text{dof} = 23.0/19$ which corresponds to a $p$-value of
2.4\%. When expressed in terms of an energy-independent normalization
factor $f$, the best fit is obtained at
\begin{equation}
  f = 0.935 \pm 0.024 \,,\quad
  \chi^2_\text{min}/\text{dof} = 15.7/18 ~(p = 61\%) \,, \quad
  \Delta\chi^2_{f=1} = 7.25 \quad (2.7\sigma) \,.
\end{equation}
Here $\Delta\chi^2_{f=1}$ denotes the improvement in $\chi^2$ compared
to a fit with $f = 1$. Clearly the $p$-value increases drastically
when $f$ is allowed to float, leading to a preference for $f \neq 1$
at the $2.7\sigma$ confidence level.  This is our result for the
significance of the reactor anomaly. Let us mention that (obviously)
this result depends on the assumed systematic errors. While we have no
particular reason to doubt any of the quoted errors, we have checked
that when an adhoc additional normalization uncertainty of 2\% (3\%)
is added, the significance is reduced to $2.1\sigma$
($1.7\sigma$). This shows that the reactor anomaly relies on the
control of systematic errors at the percent level.

\begin{table}[t]
  \centering
  \fitnumber{
    \begin{tabular}{lcccc}
      \hline\hline
      & $\sin^22\theta_{14}$ & $\Dmq_{41} \, [\eVq]$ &
      $\chi^2_\text{min}$/dof (GOF) & $\Delta\chi^2_\text{no-osc}$/dof (CL) \\
      \hline
      SBL rates only            & 0.13 & 0.44 & 11.5/17 (83\%) & 11.4/2 (99.7\%) \\
      SBL incl.\ Bugey3 spectr. & 0.10 & 1.75 & 58.3/74 (91\%) & 9.0/2  (98.9\%) \\
      SBL + Gallium             & 0.11 & 1.80 & 64.0/78 (87\%) & 14.0/2 (99.9\%) \\
      SBL + LBL                 & 0.09 & 1.78 & 93.0/113 (92\%) & 9.2/2 (99.0\%) \\
      global $\nu_e$ disapp.\   & 0.09 & 1.78 & 403.3/427 (79\%) & 12.6/2 (99.8\%) \\
      \hline\hline
  \end{tabular}}
  \mycaption{Best fit oscillation parameters and $\chi^2_\text{min}$
    values as well as $\Delta\chi^2_\text{no-osc} \equiv
    \chi^2_\text{no-osc} - \chi^2_\text{min}$ within a 3+1
    framework. Except in the row labeled ``SBL rates only'', we always
    include spectral data from Bugey3. The row ``global $\nu_e$
    disapp.'' includes the data from reactor experiments (see
    Tab.~\ref{tab:react-data}) as well as Gallium data, solar
    neutrinos and the LSND/KARMEN $\nu_e$ disappearance data from
    $\nu_e$--$\iso{C}{12}$ scattering.  The CL for the exclusion of
    the no oscillation hypothesis is calculated assuming 2~degrees of
    freedom ($|U_{e4}|$ and $\Dmq_{41}$).\label{tab:react-bfp}}
\end{table}

\begin{figure}[t]
  \centering
  \includegraphics[height=6cm]{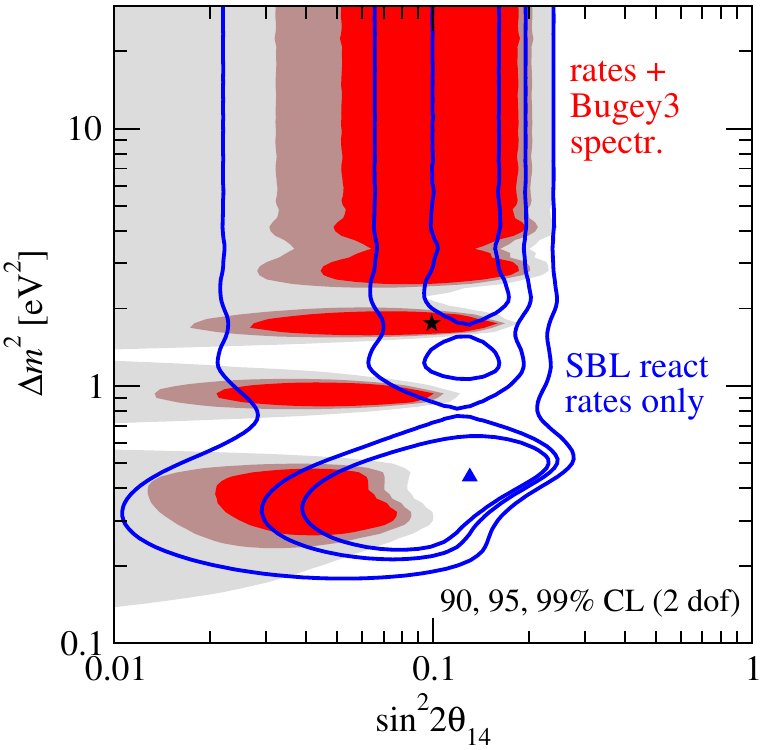} \quad
  \includegraphics[height=6cm]{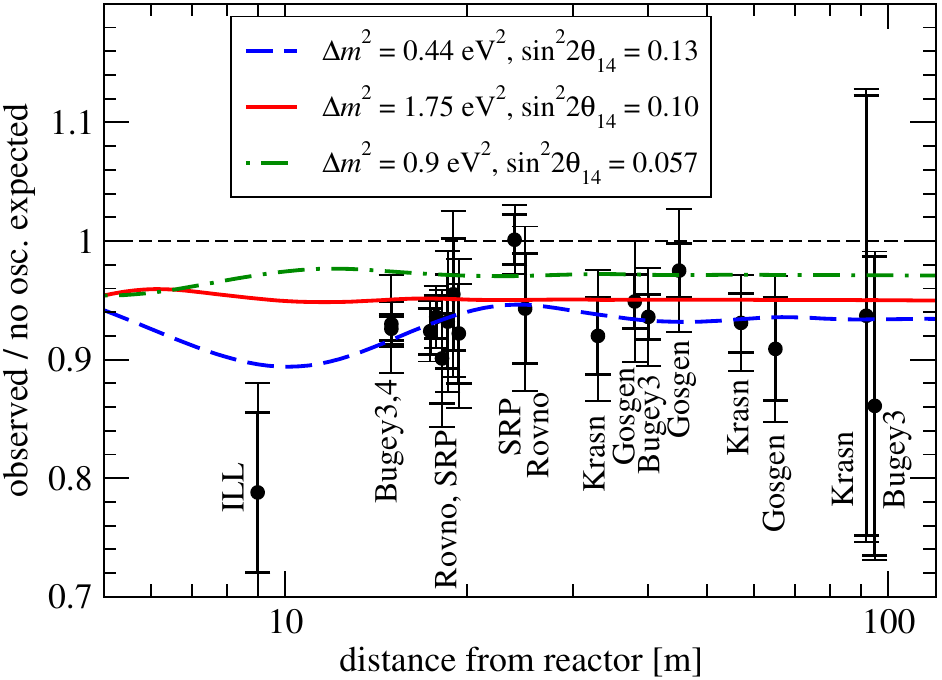}
  \mycaption{Left: Allowed regions of oscillation parameters from SBL
    reactor data in the 3+1 scheme for a rates only analysis
    (contours) as well as a fit including Bugey3 spectral data
    (colored regions). Right: Event rates in SBL reactor experiments
    compared to the predictions for three representative sets of
    oscillation parameters. The thick (thin) error bars correspond to
    uncorrelated (total) experimental errors. The neutrino flux
    uncertainty is not included in the error bars. The Rovno and SRP
    data points at 18~m have been shifted for better
    visibility.\label{fig:sbl-react}}
\end{figure}

The flux reduction suggested by the reactor anomaly can be explained
by sterile neutrino oscillations. In Tab.~\ref{tab:react-bfp} we give
the best fit points and $\chi^2$ values obtained by fitting SBL
reactor data in a 3+1 framework. The allowed regions in $\Dmq_{41}$
and $\sin^22\theta_{14}$ are shown in Fig.~\ref{fig:sbl-react} (left)
for a rate-only analysis as well as a fit including also Bugey3
spectral data. Both analyses give consistent results, with the main
difference being that the spectral data disfavors certain values of
$\Dmq_{41}$ around $0.6-0.7~\eVq$ and 1.3~\eVq. The right panel of
Fig.~\ref{fig:sbl-react} shows the predicted rate suppression as a
function of the baseline compared to the data. We show the prediction
for the two best fit points from the left panel as well as one point
located in the island around $\Dmq_{41} \simeq 0.9~\eVq$, which will
be important in the combined fit with SBL appearance data. We observe
that for the rate-only best fit point with $\Dmq_{41} = 0.44~\eVq$ the
prediction follows the tendency suggested by the ILL, Bugey4, and SRP
(24~km) data points. This feature is no longer present for $\Dmq_{41}
\gtrsim 1~\eVq$, somewhat preferred by Bugey3 spectral data, where
oscillations happen at even shorter baselines. However, from the GOF
values given in Tab.~\ref{tab:react-bfp} we conclude that also those
solutions provide a good fit to the data.

%------------------------------------------------------------------------------
\subsection{The Gallium anomaly}
%------------------------------------------------------------------------------

The response of Gallium solar neutrino experiments has been tested by
deploying radioactive $^{51}$Cr or $^{37}$Ar sources in the
GALLEX~\cite{Hampel:1997fc, Kaether:2010ag} and
SAGE~\cite{Abdurashitov:1998ne, Abdurashitov:2005tb} detectors.
Results are reported as ratios of observed to expected rates, where
the latter are traditionally computed using the best fit cross section
from Bahcall~\cite{Bahcall:1997eg}, see
\textit{e.g.}~\cite{Giunti:2010zu}. The values for the cross sections
weighted over the 4 (2) neutrino energy lines from Cr (Ar)
from~\cite{Bahcall:1997eg} are $\sigma_\text{B}(\text{Cr}) =
58.1\times 10^{-46}\,\text{cm}^2$, $\sigma_\text{B}(\text{Ar}) =
70.0\times 10^{-46}\,\text{cm}^2$. While the cross section for
$\iso{Ga}{71} \to \iso{Ge}{71}$ into the ground state of
$\iso{Ge}{71}$ is well known from the inverse reaction there are large
uncertainties when the process proceeds via excited states of
$^{71}$Ge at 175 and 500~keV. Following~\cite{Bahcall:1997eg}, the
total cross section can be written as
\begin{equation}\label{eq:CS-Gal}
  \sigma(X) = \sigma_\text{g.s.}(X) \left( 1 +
  a_X \frac{\text{BGT}_{175}}{\text{BGT}_\text{g.s.}} +
  b_X \frac{\text{BGT}_{500}}{\text{BGT}_\text{g.s.}}\right)
\end{equation}
with $X = \text{Cr}$, $\text{Ar}$. The coefficients $a_\text{Cr} =
0.669$, $b_\text{Cr} = 0.220$, $a_\text{Ar} = 0.695$, $b_\text{Ar} =
0.263$ are phase space factors. The ground state cross sections are
precisely known~\cite{Bahcall:1997eg}: $\sigma_\text{g.s.}(\text{Cr})
= 55.2\times 10^{-46} \,\text{cm}^2$, $\sigma_\text{g.s.}(\text{Ar}) =
66.2\times 10^{-46} \,\text{cm}^2$.  BGT denote the Gamov-Teller
strength of the transitions, which have been determined recently by
dedicated measurements~\cite{Frekers:2011zz} as
\begin{equation}\label{eq:BGT}
  \frac{\text{BGT}_{175}}{\text{BGT}_\text{g.s.}} = 0.0399 \pm 0.0305\,,\qquad
  \frac{\text{BGT}_{500}}{\text{BGT}_\text{g.s.}} = 0.207  \pm 0.016 \,.
\end{equation}
In our analysis we use these values together with
Eq.~\eqref{eq:CS-Gal} for the cross section.

This means that the ratios of observed to expected rates based on the
Bahcall prediction have to be rescaled by a factor 0.982 (0.977) for
the Cr (Ar) experiments, so that we obtain for them the following
updated numbers for our fits:
\begin{equation} \label{eq:Gal-ratios}
  \text{GALLEX:}\left\lbrace
  \begin{aligned}
    R_1(\text{Cr}) &= 0.94 \pm 0.11~\text{\cite{Kaether:2010ag}}\\
    R_2(\text{Cr}) &= 0.80 \pm 0.10~\text{\cite{Kaether:2010ag}}
  \end{aligned}
  \right. \,,
  \quad
  \text{SAGE:} \left\lbrace
  \begin{aligned}
    R_3(\text{Cr}) &= 0.93 \pm 0.12~\text{\cite{Abdurashitov:1998ne}}\\
    R_4(\text{Ar}) &= 0.77 \pm 0.08~\text{\cite{Abdurashitov:2005tb}}
  \end{aligned}
  \right. \,.
\end{equation}
Here, we have symmetrized the errors, and we have included only
experimental errors, but not the uncertainty on the cross section (see
below).

We build a $\chi^2$ out of the four data points from GALLEX and SAGE
and introduce two pulls corresponding to the systematic uncertainty of
the two transitions to excited state according to
Eq.~\eqref{eq:BGT}. The determination of BGT$_{175}$ is relatively
poor, with zero being allowed at $2\sigma$. In order to avoid
unphysical negative contributions from the 175~keV state, we restrict
the domain of the corresponding pull parameter accordingly.  Fitting
the four data points with a constant neutrino flux normalization
factor $r$ we find
\begin{equation}
  \chi^2_\text{min} = 2.26 / 3 \, \text{dof} \,,
  \qquad
  r_\text{min} = 0.84^{+0.054}_{-0.051} \,,
  \qquad
  \Delta\chi^2_{r=1} = 8.72 \quad (2.95\sigma)
\end{equation}
Because of the different cross sections used, these results differ
from the ones obtained in~\cite{Giunti:2010zu}, where the best fit
point is at $r = 0.76$, while the significance is comparable, around
$3\sigma$. An updated analysis including also a discussion of the
implications of the measurement in~\cite{Frekers:2011zz} can be found
in~\cite{Giunti:2012tn}.

\begin{figure}[t]
  \centering
  \includegraphics[width=0.65\textwidth]{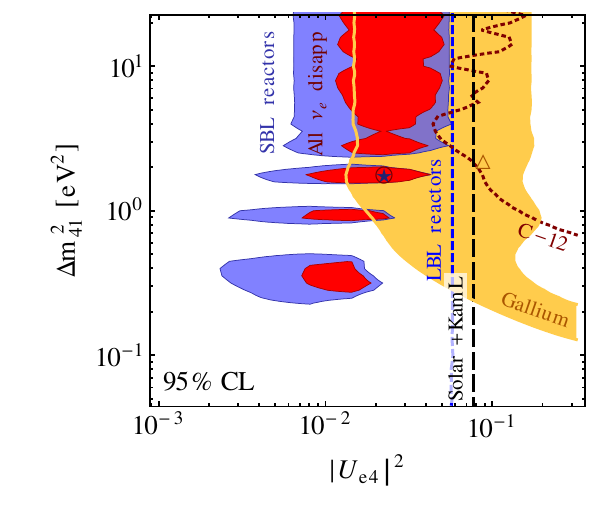}
  \mycaption{Allowed regions at 95\%~CL (2 dof) for 3+1
    oscillations. We show SBL reactor data (blue shaded), Gallium
    radioactive source data (orange shaded), $\nu_e$ disappearance
    constraints from $\nu_e$--\iso{C}{12} scattering data from LSND
    and KARMEN (dark red dotted), long-baseline reactor data from
    CHOOZ, Palo Verde, DoubleChooz, Daya Bay and RENO (blue
    short-dashed) and solar+KamLAND data (black long-dashed). The red
    shaded region is the combined region from all these $\nu_e$ and
    $\bar\nu_e$ disappearance data sets.\label{fig:nue-global}}
\end{figure}

The event deficit in radioactive source experiments can be explained
by assuming $\nu_e$ mixing with an eV-scale state, such that $\nu_e$
disappearance happens within the detector
volume~\cite{Acero:2007su}. We fit the Gallium data in the 3+1
framework by averaging the oscillation probability over the detector
volume using the geometries given in~\cite{Acero:2007su}. The
resulting allowed region at 95\% confidence level is shown in orange
in Fig.~\ref{fig:nue-global}. Consistent with the above discussion we
find mixing angles somewhat smaller than those obtained by the authors
of~\cite{Giunti:2010zu}.  The best fit point from combined Gallium+SBL
reactor data is given in Tab.~\ref{tab:react-bfp}, and the
no-oscillation hypothesis is disfavored at 99.9\%~CL (2 dof) or
$3.3\sigma$ compared to the 3+1 best fit point.

\begin{table}[t]
  \centering
  \begin{tabular}{lccccccc}
    \hline\hline
    & $\Dmq_{41}$& $\Dmq_{51}$ & $\theta_{14}$ & $\theta_{15}$
    & $\chi^2_\text{min}$ (GOF) & $\Delta\chi^2_{3+1}$ (CL) & $\Delta\chi^2_\text{no-osc}$ (CL)
    \\
    \hline
    SBLR & 0.46 & 0.87 & 0.12 & 0.13 & 53.0/(76-4) (95\%) & 5.3 (93\%) &
    14.3 (99.3\%)
    \\
    SBLR+gal & 0.46 & 0.87 & 0.12 & 0.14 & 60.2/(80-4) (90\%) & 3.8 (85\%) &
    17.8 (99.9\%)
    \\
    \hline\hline
  \end{tabular}
  \mycaption{Best fit point of SBL reactor data and SBL reactor +
    Gallium data in a 3+2 oscillation scheme. We give the mass-squared
    differences in \eVq\ and the mixing angles in radians. The
    relation to the mixing matrix elements is $|U_{e4}| =
    \cos\theta_{15}\sin\theta_{14}$ and $|U_{e5}| =
    \sin\theta_{15}$. The $\Delta\chi^2$ relative to 3+1 oscillations
    is evaluated for 2 dof, corresponding to the two additional
    parameters, while for the $\Delta\chi^2$ relative to
    no-oscillations we use 4 dof.\label{tab:3+2react}}
\end{table}

Let us consider now the Gallium and SBL reactor data in the framework
of two sterile neutrinos, in particular in the 3+2 scheme. SBL $\nu_e$
and $\bar\nu_e$ disappearance data depend on 4 parameters in this
case, $\Dmq_{41}$, $\Dmq_{51}$, and the two mixing angles
$\theta_{14}$ and $\theta_{15}$ (or, equivalently, the moduli of the
two matrix elements $U_{e4}$ and $U_{e5}$). We report the best fit
points from SBL reactor data and from SBL reactor data combined with
the Gallium source data in Tab.~\ref{tab:3+2react}.  For these two
cases we find an improvement of 5.3 and 3.8 units in $\chi^2$,
respectively, when going from the 3+1 scenario to the 3+2
case. Considering that the 3+2 model has two additional parameters
compared to 3+1, we conclude that there is no improvement of the fit
beyond the one expected by increasing the number of parameters, and
that SBL $\parenbar\nu_e$ data sets show no significant preference for
3+2 over 3+1. This is also visible from the fact that the confidence
level at which the no oscillation hypothesis is excluded does not
increase for 3+2 compared to 3+1, see the last columns of
Tabs.~\ref{tab:react-bfp} and~\ref{tab:3+2react}. There the
$\Delta\chi^2$ is translated into a confidence level by taking into
account the number of parameters relevant in each model,
\textit{i.e.}, 2 for 3+1 and 4 for 3+2.

%------------------------------------------------------------------------------
\subsection{Global data on $\nu_e$ and $\bar\nu_e$ disappearance}
%------------------------------------------------------------------------------

Let us now consider the global picture regarding $\pbn_e$
disappearance. In addition to the short-baseline reactor and Gallium
data discussed above, we now add data from the following experiments:
\begin{itemize}
\item The remaining reactor experiments at a long baseline (``LBL
  reactors'') and the very long-baseline reactor experiment KamLAND,
  see table~\ref{tab:react-data}.

\item Global data on solar neutrinos, see appendix~\ref{app:solar} for
  details.

\item LSND and KARMEN measurements of the reaction $\nu_e +
  \iso{C}{12} \to e^- + \iso{N}{12}$~\cite{Auerbach:2001hz,
    Armbruster:1998uk}. The experiments have found agreement with the
  expected cross section~\cite{Fukugita:1988hg}, hence their
  measurements constrain the disappearance of $\nu_e$ with eV-scale
  mass-squared differences~\cite{Reichenbacher:2005nc, Conrad:2011ce}.
  Details on our analysis of the $\iso{C}{12}$ scattering data are
  given in appendix~\ref{app:carbon}.
\end{itemize}

So far the LBL experiments DayaBay and RENO have released only data on
the relative comparison of near ($L\sim 400$~m) and far ($L\sim
1.5$~km) detectors, but no information on the absolute flux
determination is available. Therefore, their published data are
essentially insensitive to oscillations with eV-scale neutrinos and
they contribute only indirectly via constraining $\theta_{13}$. In our
analysis we include a free, independent flux normalization factor for
each of those two experiments.  Chooz and DoubleChooz both lack a near
detector. Therefore, in the official analyses performed by the
respective collaborations the Bugey4 measurement is used to normalize
the flux. This makes the official Chooz and DoubleChooz results on
$\theta_{13}$ also largely independent of the presence of sterile
neutrinos. However, the absolute rate of Bugey4 in terms of the flux
predictions is published (see Tab.~\ref{tab:react-data}) and we can
use this number to obtain an absolute flux prediction for Chooz and
DoubleChooz. Therefore, in our analysis Chooz and DoubleChooz (as well
as Palo Verde) by themselves also show some sensitivity to sterile
neutrino oscillations. In a combined analysis of Chooz and DoubleChooz
with SBLR data the official analyses are recovered
approximately. Previous considerations of LBL reactor experiments in
the context of sterile neutrinos can be found in
Refs.~\cite{Bandyopadhyay:2007rj, Bora:2012pi, Giunti:2011vc,
  Bhattacharya:2011ee}.

We show in Tab.~\ref{tab:react-bfp} a combined analysis of the SBL and
LBL reactor experiments (row denoted by ``SBL+LBL''), where we
minimize with respect to $\theta_{13}$. We find that the significance
of the reactor anomaly is not affected by the inclusion of LBL
experiments and finite $\theta_{13}$. The $\Delta\chi^2_\text{no-osc}$
even slightly increases from 9.0 to 9.2 when adding LBL data to the
SBL data (``no-osc'' refers here to $\theta_{14}=0$). Hence, we do not
agree with the conclusions of Ref.~\cite{Zhang:2013ela}, which finds
that the significance of the reactor anomaly is reduced to $1.4\sigma$
when LBL data and a finite value of $\theta_{13}$ is taken into
account.

Solar neutrinos are also sensitive to sterile neutrino mixing (see
\textit{e.g.}~\cite{Giunti:2009xz, Palazzo:2011rj,
  Palazzo:2012yf}). The main effect of the presence of $\nu_e$ mixing
with eV states is an over-all flux reduction.  While this effect is
largely degenerate with $\theta_{13}$, a non-trivial bound is obtained
in the combination with DayaBay, RENO and KamLAND.  KamLAND is
sensitive to oscillations driven by $\Dmq_{21}$ and $\theta_{12}$,
whereas sterile neutrinos affect the overall normalization, degenerate
with $\theta_{13}$.  The matter effect in the sun as well as SNO NC
data provide additional signatures of sterile neutrinos, beyond an
overall normalization. As we will show in Sec.~\ref{sec:numu-dis}
solar data depend also on the mixing angles $\theta_{24}$ and
$\theta_{34}$, controlling the fraction of $\nu_e\to\nu_s$
transitions, see \textit{e.g.}~\cite{Giunti:2009xz}. As discussed in
appendix~\ref{app:solar}, in the limit $\Dmq_{i1} = \infty$ for $i \ge
3$, solar data depends on 6 real mixing parameters, 1 complex phase
and $\Dmq_{21}$. Hence, in a 3+1 scheme all six mixing angles are
necessary to describe solar data in full generality. However, once
other constraints on mixing angles are taken into account the effect
of $\theta_{24}$, $\theta_{34}$, and the complex phase are tiny and
numerically have a negligible impact on our results. Therefore we set
$\theta_{24} = \theta_{34} = 0$ for the solar neutrino analysis in
this section. In this limit solar data becomes also independent of the
complex phase.

The results of our fit to global $\pbn_e$ disappearance data are shown
in Fig.~\ref{fig:nue-global} and the best fit point is given in
Tab.~\ref{tab:react-bfp}. For this analysis the mass-squared
differences $\Dmq_{21}$ and $\Dmq_{31}$ have been fixed, whereas we
marginalize over the mixing angles $\theta_{12}$ and $\theta_{13}$. We
see from Fig.~\ref{fig:nue-global} that the parameter region favored
by short-baseline reactor and Gallium data is well consistent with
constraints from long-baseline reactors, KARMEN's and LSND's $\nu_e$
rate, and with solar and KamLAND data.

Recently, data from the Mainz~\cite{Kraus:2012he} and
Troitsk~\cite{Belesev:2012hx} tritium beta-decay experiments have been
re-analyzed to set limits on the mixing of $\nu_e$ with new
$\gtrsim$~eV neutrino mass states. Taking the results
of~\cite{Belesev:2012hx} at face value, the Troitsk limit would
cut-off the high-mass region in Fig.~\ref{fig:nue-global} at around
100~\eVq~\cite{Giunti:2012bc} (above the plot-range shown in the
figure). The bounds obtained in~\cite{Kraus:2012he} are somewhat
weaker. The differences between the limits obtained
in~\cite{Kraus:2012he} and~\cite{Belesev:2012hx} depend on assumptions
concerning systematic uncertainties and therefore we prefer not to
explicitly include them in our fit. The sensitivity of future tritium
decay data from the KATRIN experiment has been estimated
in~\cite{Riis:2010zm}. Implications of sterile neutrinos for
neutrino-less double beta-decay have been discussed recently
in~\cite{Li:2011ss, Barry:2011wb, Giunti:2011cp}.

\begin{figure}[t]
  \centering
  \includegraphics[width=0.49\textwidth]{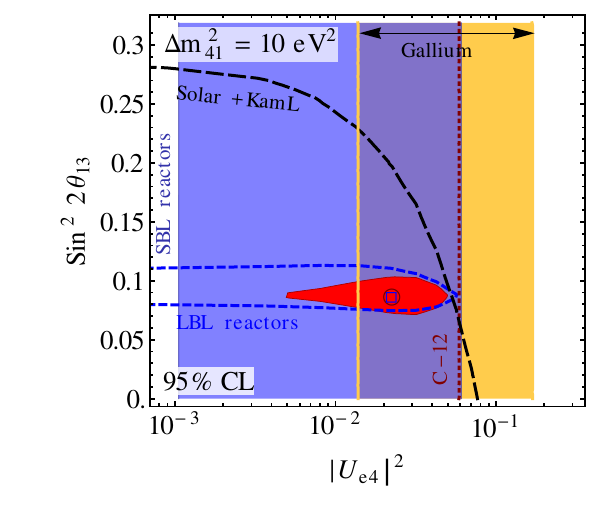}
  \includegraphics[width=0.49\textwidth]{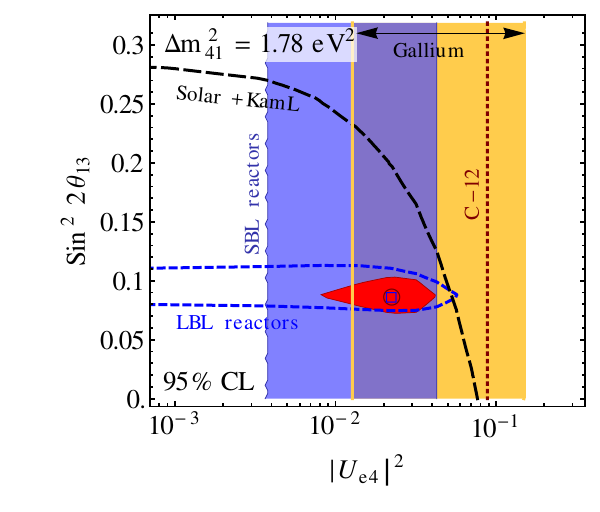}
  \mycaption{Constraints on $\nu_e$ and $\bar\nu_e$ disappearance in a
    $3+1$ model at two different fixed values of $\Dmq_{41}$. Regions
    are shown at 95\%~CL (2~dof) with respect to the minimum $\chi^2$
    at the fixed $\Dmq_{41}$.  We show constraints from the radio
    chemical Gallium experiments using radioactive sources (orange
    band), from short-baseline reactor experiments (blue band), from
    the KARMEN and LSND measurements of the $\nu_e$--$\iso{C}{12}$
    cross section (dark red dotted line), from long-baseline reactor
    experiments (blue dashed line), from the combined solar+KamLAND
    data (black dashed line), and from the the combination of all
    aforementioned experiments (red shaded
    region).\label{fig:th13-th14}}
\end{figure}

Let us now address the question whether the presence of a sterile
neutrino affects the determination of the mixing angle $\theta_{13}$
(see also~\cite{Bhattacharya:2011ee, Zhang:2013ela}).  In
Fig.~\ref{fig:th13-th14} we show the combined determination of
$\theta_{13}$ and $\theta_{14}$ for two fixed values of
$\Dmq_{41}$. The left panel corresponds to a relatively large value of
10~\eVq, whereas for the right panel we have chosen the value favored
by the global $\parenbar{\nu}_e$ disappearance best fit point,
\fitnumber{1.78~\eVq}.  The mass-squared differences $\Dmq_{21}$ and
$\Dmq_{31}$ have been fixed, whereas we marginalize over the mixing
angle $\theta_{12}$. We observe a clear complementarity of the
different data sets: SBL reactor and Gallium data determine
$|U_{e4}|$, since oscillations are possible only via $\Dmq_{41}$, all
other mass-squared differences are effectively zero for them.  For LBL
reactors $\Dmq_{41}$ can be set to infinity, $\Dmq_{31}$ is finite,
and $\Dmq_{21}$ is effectively zero; therefore they provide an
unambiguous determination of $\theta_{13}$ by comparing near and far
detector data. The upper bound on $|U_{e4}|$ from LBL reactors is
provided by Chooz, Palo Verde, DoubleChooz, since for those
experiments also information on the absolute flux normalization can be
used, as mentioned above.  In contrast, for solar neutrinos and
KamLAND, both $\Dmq_{41}$ and $\Dmq_{31}$ are effectively infinite,
and $\theta_{13}$ and $\theta_{14}$ affect essentially the overall
normalization and are largely degenerate, as visible the figure.

In conclusion, the $\theta_{13}$ determination is rather stable with
respect to the presence of sterile neutrinos. We note, however, that
its interpretation becomes slightly more complicated. For instance, in
the 3+1 scheme using the parametrization from Tab.~\ref{tab:counting},
the relation between mixing matrix elements and mixing angles is
$|U_{e3}| = \cos\theta_{14} \sin\theta_{13}$ and $|U_{e4}| =
\sin\theta_{14}$. Hence, the one-to-one correspondence between
$|U_{e3}|$ and $\theta_{13}$ as in the three-flavor case is spoiled.

% =============================================================================
\section{$\nu_\mu$, $\bar\nu_\mu$, and neutral-current disappearance searches}
\label{sec:numu-dis}
% =============================================================================

In this section we discuss the constraints on the mixing of
$\parenbar\nu_\mu$ and $\parenbar\nu_\tau$ with new eV-scale mass
states. In the 3+1 scheme this is parametrized by $|U_{\mu 4}|$ and
$|U_{\tau 4}|$, respectively.  In terms of the mixing angles as
defined in Eq.~\eqref{eq:rotations} we have $|U_{\mu 4}| =
\cos\theta_{14}\sin\theta_{24}$ and $|U_{\tau 4}| =
\cos\theta_{14}\cos\theta_{24}\sin\theta_{34}$.  In the present paper
we include data sets from the following experiments to constrain
$\parenbar\nu_\mu$ and $\parenbar\nu_\tau$ mixing with eV states:
\begin{itemize}
  \item SBL $\nu_\mu$ disappearance data from
    CDHS~\cite{Dydak:1983zq}. Details of our simulation are given
    in~\cite{Grimus:2001mn}.

  \item Super-Kamiokande. It has been pointed out
    in~\cite{Bilenky:1999ny} that atmospheric neutrino data from
    Super-Kamiokande provide a bound on the mixing of $\nu_\mu$ with
    eV-scale mass states, \textit{i.e.}, on the mixing matrix elements
    $|U_{\mu 4}|$, $|U_{\mu 5}|$. In addition, neutral-current matter
    effects provide a constraint on $|U_{\tau 4}|$, $|U_{\tau 5}|$. A
    discussion of the effect is given in the appendix
    of~\cite{Maltoni:2007zf}. Details on our analysis and references
    are given in appendix~\ref{app:atm}.

  \item MiniBooNE~\cite{AguilarArevalo:2009yj, Cheng:2012yy}.  Apart
    from the $\parenbar{\nu}_e$ appearance search, MiniBooNE can also
    look for SBL $\parenbar\nu_\mu$ disappearance.  Details on our
    analysis are given in appendix~\ref{app:MB}.

  \item MINOS~\cite{Adamson:2010wi, Adamson:2011ku}. The MINOS
    long-baseline experiment has published data on charged current
    (CC) $\parenbar{\nu}_\mu$ disappearance as well as on the neutral
    current (NC) count rate. Both are based on a comparison of near
    and far detector measurements. In addition to providing the most
    precise determination of $\Dmq_{31}$ (from CC data), those data
    can also be used to constrain sterile neutrino mixing, where CC
    (NC) data are mainly relevant for $|U_{\mu 4}|$, $|U_{\mu 5}|$
    ($|U_{\tau 4}|$, $|U_{\tau 5}|$). See appendix~\ref{app:minos} for
    details.
\end{itemize}
Additional constrains on $\nu_\mu$ mixing with eV-scale states (not
used in this analysis) can be obtained from data from the Ice Cube
neutrino telescope~\cite{Nunokawa:2003ep, Choubey:2007ji,
  Razzaque:2011ab, Barger:2011rc, Razzaque:2012tp, Esmaili:2012nz}.

\begin{figure}[t]
  \centering
  \raisebox{-5mm}{\includegraphics[width=0.5\textwidth]{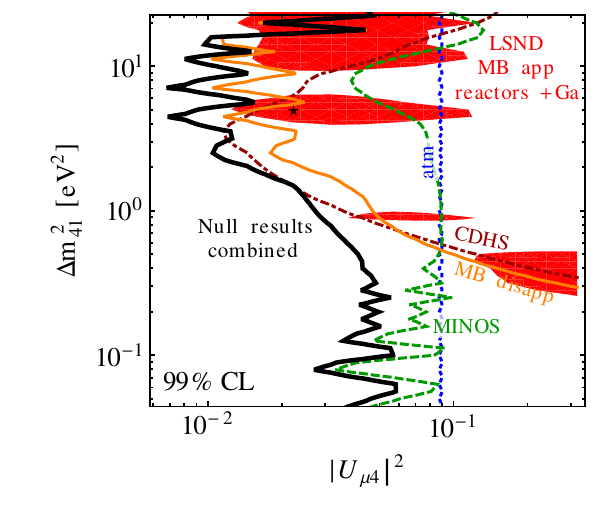}}
  \includegraphics[width=0.44\textwidth]{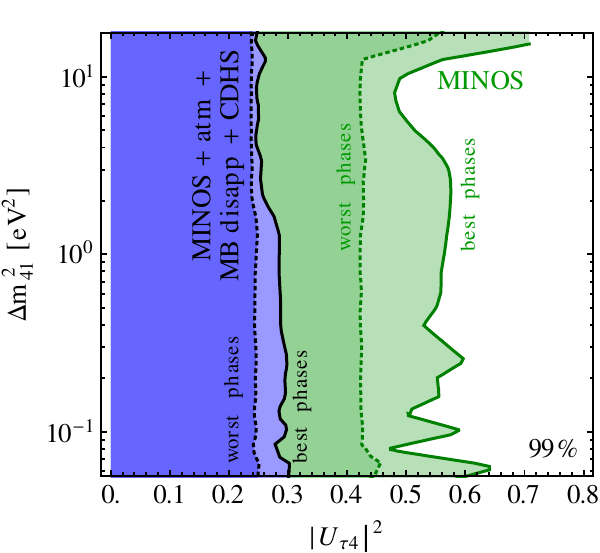}
  \mycaption{Left: Constraints in the plane of $|U_{\mu 4}|^2$ and
    $\Dmq_{41}$ at 99\%~CL (2~dof) from CDHS, atmospheric neutrinos,
    MiniBooNE disappearance, MINOS CC and NC data, and the combination
    of them. We minimize with respect to $|U_{\tau 4}|$ and the
    complex phase $\varphi_{24}$. In red we show the region preferred
    by LSND and MiniBooNE appearance data combined with reactor and
    Gallium data on $\pbn_e$ disappearance, where for fixed $|U_{\mu
      4}|^2$ we minimize with respect to $|U_{e4}|^2$. Right:
    Constraints in the plane of $|U_{\tau 4}|^2$ and $\Dmq_{41}$ at
    99\%~CL (2~dof) from MINOS CC + NC data (green) and the combined
    global $\pbn_\mu$ and NC disappearance data (blue region, black
    curves). We minimize with respect to $|U_{\mu 4}|$ and we show the
    weakest (``best phase'') and strongest (``worst phase'') limits,
    depending on the choice of the complex phase $\varphi_{24}$. In
    both panels we minimize with respect to $\Dmq_{31}$,
    $\theta_{23}$, and we fix $\sin^22\theta_{13} = 0.092$ and
    $\theta_{14} = 0$ (except for the evidence regions in the left
    panel).\label{fig:th24dm41}}
\end{figure}

Limits on the $|U_{\mu i}|$ row of the mixing matrix come from
$\parenbar{\nu}_\mu$ disappearance experiments. In a 3+1 scheme the
$\parenbar{\nu}_\mu$ SBL disappearance probability is given by
\begin{equation}
  P_{\mu\mu}^\text{SBL,3+1} = 1 - 4 |U_{\mu 4}|^2(1 - |U_{ \mu 4}|^2)
  \sin^2\frac{\Dmq_{41} L}{4E}
  = 1 - \sin^22\theta_{\mu\mu} \sin^2\frac{\Dmq_{41} L}{4E} \,,
\end{equation}
where we have defined an effective $\parenbar \nu_\mu$ disappearance
mixing angle by
\begin{equation}\label{eq:sinq2t_mm}
  \sin^22\theta_{\mu\mu} \equiv 4 |U_{\mu 4}|^2(1 - |U_{\mu 4}|^2)\,,
\end{equation}
\textit{i.e.}, in our parametrization~\eqref{eq:rotations} the
effective mixing angle $\theta_{\mu\mu}$ depends on both $\theta_{24}$
and $\theta_{14}$.  In contrast to the $\nu_e$ disappearance searches
discussed in the previous section, experiments probing
$\parenbar{\nu}_\mu$ disappearance have not reported any hints for a
positive signal. We show the limits from the data listed above in the
left panel of Fig.~\ref{fig:th24dm41}. Note that the MINOS limit is
based on the comparison of the data in near and far detectors. For
$\Dmq_{41}\sim 10\,\eVq$ oscillation effects become relevant at the
near detector, explaining the corresponding features in the MINOS
bound around that value of $\Dmq_{41}$, whereas the features around
$\Dmq_{41}\sim 0.1\,\eVq$ emerge from oscillation effects in the far
detector. The roughly constant limit in the intermediate range
$0.5\,\eVq \lesssim \Dmq_{41} \lesssim 3\,\eVq$ corresponds to the
limit $\Dmq_{41} \approx 0 \, (\infty)$ in the near (far) detector
adopted in~\cite{Adamson:2010wi, Adamson:2011ku}. In that range the
MINOS limit on $|U_{\mu 4}|$ is comparable to the one from SuperK
atmospheric data. For $\Dmq_{41}\gtrsim 1\,\eVq$ the limit is
dominated by CDHS and MiniBooNE disappearance data.

In Fig.~\ref{fig:th24dm41} (left) we show also the region preferred by
the hints for eV-scale oscillations from LSND and MiniBooNE appearance
data (see next section) combined with reactor and Gallium data on
$\pbn_e$ disappearance. For fixed $|U_{\mu 4}|^2$ we minimize the
corresponding $\chi^2$ with respect to $|U_{e4}|^2$ to show the
projection in the plane of $|U_{\mu 4}|^2$ and $\Dmq_{41}$.  The
tension between the hints in the $\pbn_\mu \to \pbn_e$ and $\pbn_e \to
\pbn_e$ channels compared to the limits from $\pbn_\mu \to \pbn_\mu$
data is clearly visible in this plot. We will discuss this conflict in
detail in section~\ref{sec:combi}.

Limits on the mixing of $\nu_\tau$ with eV-scale states are obtained
from data involving information from NC interactions, which allow to
distinguish between $\pbn_\mu \to \pbn_\tau$ and $\pbn_\mu\to \pbn_s$
transitions.\footnote{The searches for $\nu_\tau$ appearance at
  NOMAD~\cite{Astier:2001yj} and CHORUS~\cite{Eskut:2007rn} at short
  baselines are sensitive only to specific parameter combinations like
  $|U_{\mu 4} U_{\tau 4}|$ or $|U_{e4} U_{\tau 4}|$ and therefore do
  not provide a constraint on $|U_{\tau 4}|$ by itself.}  The relevant
data samples are atmospheric and solar neutrinos (via the NC matter
effect) and MINOS NC data. Furthermore, the parameter $|U_{\tau 4}|$
controls the relative weight of the oscillation modes
$\nu_\mu\to\nu_\tau$ and $\nu_\mu\to\nu_s$ at the ``atmospheric''
scale $\Dmq_{31}$: a large value of $|U_{\tau 4}|$ implies a large
fraction of $\nu_\mu\to\nu_s$ oscillations at the $\Dmq_{31}$
scale. The limit in the plane of $|U_{\tau 4}|^2$ and $\Dmq_{41}$ is
shown in the right panel of Fig.~\ref{fig:th24dm41}.

As follows from Eq.~\eqref{eq:LBLapp} (see also
appendix~\ref{app:phases}), in the LBL approximation relevant for
MINOS NC data a complex phase enters the oscillation probabilities,
corresponding to the combination $\arg(U_{\mu 4}^* U_{\tau 4} U_{\mu
  3} U_{\tau 3}^*)$. In our calculations we take the rotation matrix
$V_{24}$ to be complex and use the phase $\varphi_{24}$ to parametrize
this phase. In Fig.~\ref{fig:th24dm41} we illustrate the impact of
this phase by showing the strongest and weakest limit obtained when
varying $\varphi_{24}$. We observe that the limit from MINOS depends
quite significantly on this phase. The different shapes of the ``best
phase'' and ``worst phase'' regions emerge from the different
properties of CC and NC data. For the weakest limit (``best phase'')
the fit uses the freedom of the term including the complex phase,
which implies that a finite value of $\theta_{24}$ (or $|U_{\mu 4}|$)
is adopted, subject to the constraint from MINOS CC data.  Therefore
the same structure as in the left panel of Fig.~\ref{fig:th24dm41}
becomes visible also in limit on $|U_{\tau 4}|$. If we force the phase
to take on a value not favored by the fit, a smaller $\chi^2$ is
obtained for $\theta_{24}$ close to zero, which implies that the phase
actually becomes unphysical. In this case CC data are not important
for the limit on $|U_{\tau 4}|$, which then is dominated by NC
data. Because of the much worse energy reconstruction for NC events
compared to CC ones, the features induced by finite values of
$\Dmq_{41}$ in either the far or near detector become to a large
extent washed out.

The global limit on $|U_{\tau 4}|$ is actually dominated by
atmospheric neutrino data and shows only a very weak dependence on the
complex phase. In our atmospheric neutrino analysis the information on
$|U_{\tau 4}|$ enters via the NC matter effect induced by the presence
of sterile neutrinos. A large value of $|U_{\tau 4}|$ would imply a
significant matter effect in $\Dmq_{31}$ driven $\pbn_\mu$
disappearance, which is not consistent with the zenith angle
distribution observed in SuperK. We find the limit
\begin{equation}
  |U_{\tau 4}|^2 \lesssim 0.2
  \quad
  \text{at}\quad 2\sigma \, (1~\text{dof})
\end{equation}
from global data, largely independent of $\Dmq_{41}$ as well as
complex phases.

\begin{figure}[t]
  \includegraphics[width=0.32\textwidth]{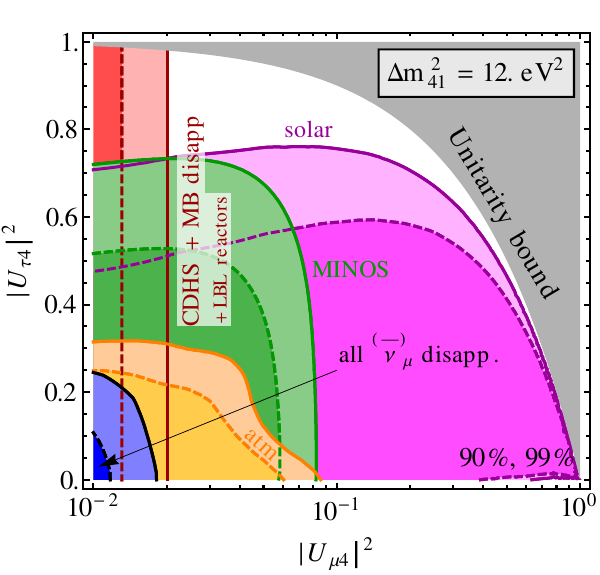}
  \includegraphics[width=0.32\textwidth]{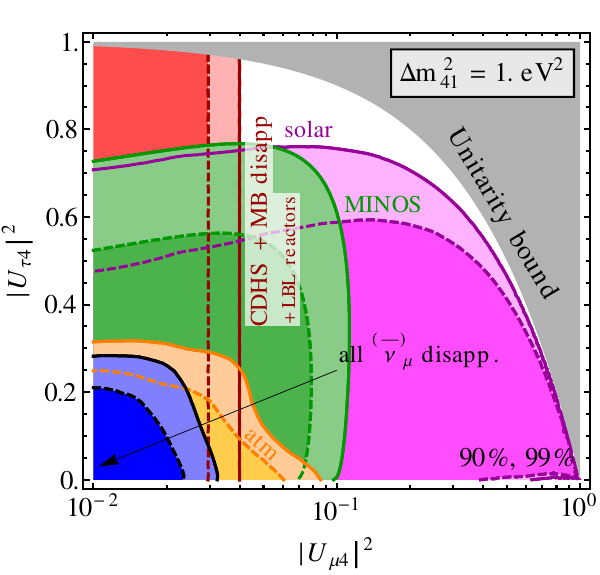}
  \includegraphics[width=0.32\textwidth]{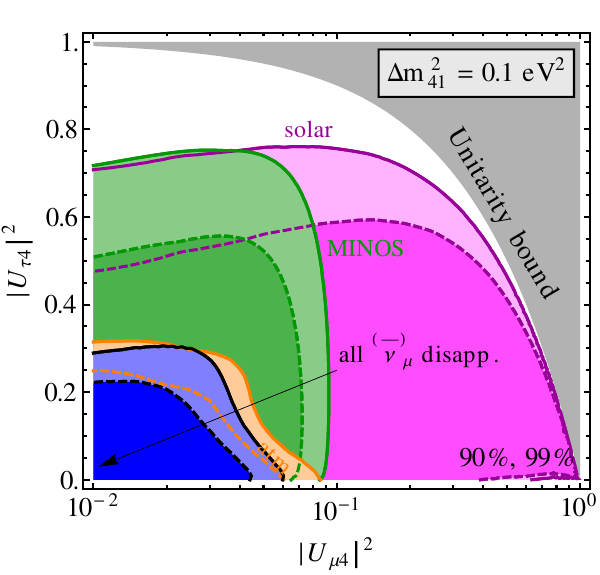}
  \mycaption{Constraints in the plane of $|U_{\mu 4}|^2$ and $|U_{\tau
      4}|^2$ for three fixed values of $\Dmq_{41}$ from MINOS CC + NC
    data (green), atmospheric neutrinos (orange), CDHS + MiniBooNE
    $\pbn_\mu$ disappearance + LBL reactors (red), and the combination
    of those data (blue). The constraint from solar neutrinos is shown
    in magenta.  Regions are shown at 90\% and 99\%~CL (2~dof) with
    respect to the $\chi^2$ minimum at the fixed $\Dmq_{41}$. We
    minimize with respect to complex phases and include effects of
    $\theta_{13}$ and $\theta_{14}$ where relevant.  The gray region
    is excluded by the unitarity requirement $|U_{\mu 4}|^2 + |U_{\tau
      4}|^2 \le 1$.  Note the different scale on the
    axes.\label{fig:th24th34}}
\end{figure}

Fig.~\ref{fig:th24th34} shows the constraints in the plane of $|U_{\mu
  4}|^2$ and $|U_{\tau 4}|^2$ for three fixed values of
$\Dmq_{41}$. We observe the comparable bound on $|U_{\mu 4}|^2$ from
MINOS (mainly CC data) and atmospheric, which however is superseded by
CDHS, MiniBooNE for $\Dmq_{41} \gtrsim 1\,\eVq$ (left and middle
panels). Those latter data however, do not provide any constraint on
$|U_{\tau 4}|^2$, where the global bound is dominated by atmospheric
neutrinos for all values of $\Dmq_{41}$ of interest. We also observe
that solar neutrinos provide a bound on $|U_{\tau 4}|^2$ of similar
strength as MINOS data, thanks to the NC matter effect and SNO NC
data.  No relevant limit can be set on $|U_{\mu 4}|^2$ from solar
neutrinos.

%==============================================================================
\section{$\nu_\mu\to\nu_e$ and $\bar\nu_\mu \to \bar\nu_e$ appearance searches}
\label{sec:nue-app}
%==============================================================================

Now we move on to the discussion of appearance searches. In contrast
to disappearance experiments which probe only one row of the mixing
matrix, \textit{i.e.}, only the elements $|U_{\alpha i}|$ for fixed
$\alpha$, an appearance experiment in the channel
$\parenbar{\nu}_\alpha \to \parenbar{\nu}_\beta$ is sensitive to two
rows via combinations like $|U_{\alpha i} U_{\beta i}|$ and
potentially to some complex phases.  In the SBL approximation the 3+1
appearance probability in the phenomenologically most relevant channel
$\parenbar{\nu}_\mu \to \parenbar{\nu}_e$ takes the form
\begin{equation}
  P_{\parenbar{\nu}_\mu \to \parenbar{\nu}_e}^\text{SBL,3+1}
  = 4 |U_{\mu 4} U_{e 4}|^2
  \sin^2\frac{\Dmq_{41} L}{4E}
  = \sin^22\theta_{\mu e} \sin^2\frac{\Dmq_{41} L}{4E} \,,
\end{equation}
where we have defined an effective mixing angle by
\begin{equation}\label{eq:sinq2t_me}
  \sin^22\theta_{\mu e} \equiv 4 |U_{\mu 4} U_{e 4}|^2 \,.
\end{equation}
In the parametrization from Eq.~\eqref{eq:rotations} we obtain $\sin
2\theta_{\mu e} = \sin\theta_{24} \sin 2\theta_{14}$.  The oscillation
probability in the 3+2 scheme is given in Eq.~\eqref{eq:SBLapp}.  The
3+1 SBL appearance probability does not depend on complex phases,
whereas in the 3+2 scheme CP violation via complex phases is possible
at SBL~\cite{Karagiorgi:2006jf, Maltoni:2007zf}.

Our analyses of LSND~\cite{Aguilar:2001ty},
KARMEN~\cite{Armbruster:2002mp}, NOMAD~\cite{Astier:2003gs}
$\parenbar\nu_\mu \to \parenbar\nu_e$ appearance data are based
on~\cite{Grimus:2001mn, PalomaresRuiz:2005vf, Maltoni:2007zf}, where
references and technical details can be found.  Our analyses of
E776~\cite{Borodovsky:1992pn} and ICARUS~\cite{Antonello:2012pq}, used
for the first time in the present paper, are described in
appendices~\ref{app:E776} and~\ref{app:icarus},
respectively.\footnote{Recently also the OPERA experiment presented
  results from a $\nu_\mu\to\nu_e$ appearance
  search~\cite{Agafonova:2013xsk}. The obtained limit is comparable to
  the one from ICARUS~\cite{Antonello:2012pq}.} In the case of LSND,
we use only the decay-at-rest (DAR) data which are most sensitive to
oscillations.  Decay-in-flight (DIF) data on $\nu_\mu\to\nu_e$ are
consistent with the signal seen in DAR data, however the significance
of the oscillation signal for DIF is much less than for DAR. A
combined DAR-DIF analysis in a two-neutrino framework would shift the
allowed region to somewhat smaller values of the mixing angle. A
detailed discussion of LSND DAR versus DIF in the context of 3+1
neutrino oscillations can be found in~\cite{Maltoni:2002xd}.

\begin{figure}[t]
  \centering
  \includegraphics[width=\textwidth]{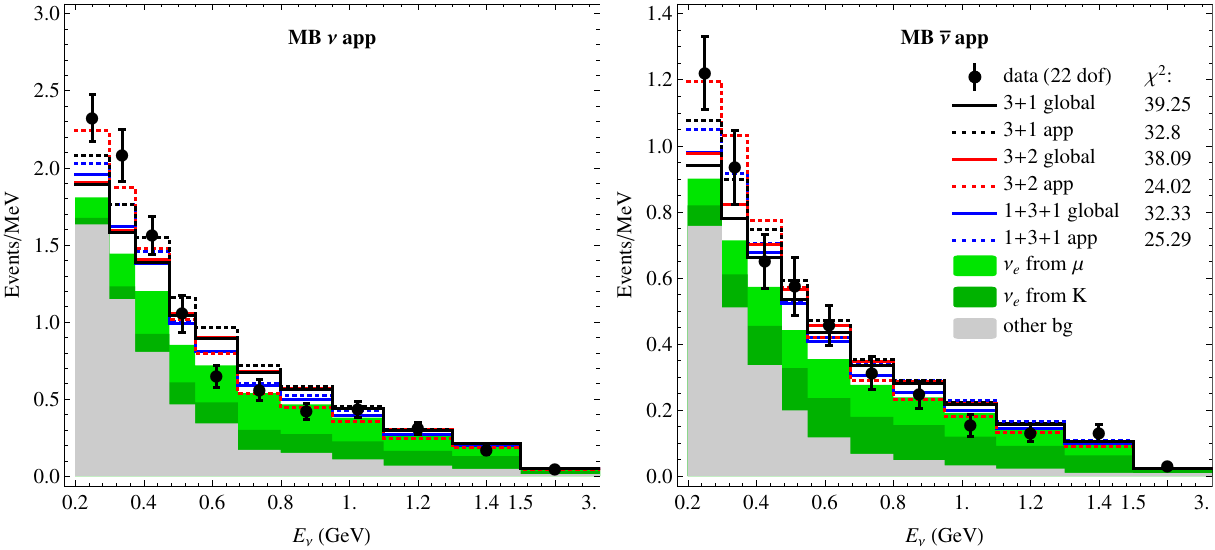}
  \mycaption{MiniBooNE neutrino (left) and anti-neutrino (right) data
    compared to the predicted spectra for the 3+1, 3+2, and 1+3+1 best
    fit points for the combined appearance data (the data set used in
    Fig.~\ref{fig:app}) and global data including
    disappearance. Shaded histograms correspond to the unoscillated
    backgrounds. The predicted spectra include the effect of
    background oscillations. The corresponding $\chi^2$ values (for
    combined neutrino and anti-neutrino data) are also given in the
    plot.\label{fig:mb-spectrum}}
\end{figure}

In our analysis of the MiniBooNE $\nu_e$ and $\bar\nu_e$ appearance
search we use the latest data\footnote{The recent updated analysis
  from MiniBooNE~\cite{Aguilar-Arevalo:2013ara} is based on the same
  data as~\cite{AguilarArevalo:2012va}, corresponding to $6.46\times
  10^{20}$ protons on target in neutrino mode and $11.27\times
  10^{20}$ protons on target in anti-neutrino mode.}
from~\cite{AguilarArevalo:2012va}, following closely the analysis
instructions provided by the collaboration.  Details are given in
appendix~\ref{app:MB}. Since their very first data release in
2007~\cite{AguilarArevalo:2007it}, MiniBooNE observe an excess of
events over expected background in the low energy ($\lesssim 500$~MeV)
region of the event spectrum~\cite{AguilarArevalo:2008rc}. Since the
spectral shape of the excess is difficult to explain in a two-flavor
oscillation framework, historically the analysis window has been
(somewhat artificially) divided into a low energy region containing
the excess events and a high energy part with no excess.\footnote{The
  importance of energy reconstruction effects for the low energy
  excess has been pointed out in Refs.~\cite{Martini:2012fa,
    Martini:2012uc}, see also~\cite{Aguilar-Arevalo:2013ara}.}
Preliminary results from anti-neutrinos showed also some indication
for an event excess in the high energy part of the
spectrum~\cite{AguilarArevalo:2010wv} which indicated the need for CP
violation in order to reconcile neutrino and anti-neutrino data.
However, for the most recent data~\cite{AguilarArevalo:2012va,
  MBnu2012} the shapes of the neutrino and anti-neutrino spectra
appear to be consistent with each other, showing excess events below
around 500~MeV and data consistent with background in the high energy
region, see Fig.~\ref{fig:mb-spectrum}. In our work we always analyse
the full energy spectrum for both neutrinos and
anti-neutrinos. Contrary to the analysis of the MiniBooNE
collaboration we take into account oscillations of all background
components in a consistent way, according to the particular
oscillation framework to be tested, see appendix~\ref{app:MB} for
details.

\begin{figure}[t]
  \centering
  \includegraphics[width=0.5\textwidth]{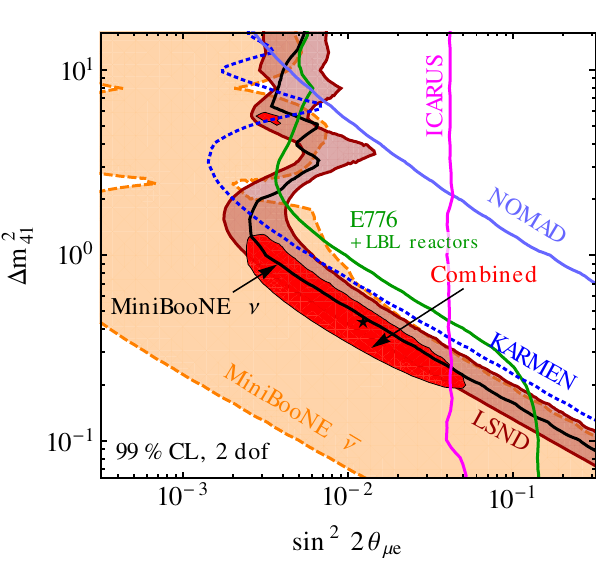}
  \mycaption{Allowed regions and upper bounds at 99\%~CL (2~dof) for
    $\parenbar{\nu}_\mu \to \parenbar{\nu}_e$ appearance experiments
    in the 3+1 scheme. We show the regions from LSND and MiniBooNE
    anti-neutrino data and the bounds from MiniBooNE neutrinos,
    KARMEN, NOMAD, ICARUS, and E776. The latter is combined with LBL
    reactor data in order to constrain the oscillations of the
    $\parenbar{\nu}_e$ backgrounds; this leads to a non-vanishing
    bound on $\sin^22\theta_{\mu e}$ from E776 at low $\Delta
    m^2_{41}$. The red region corresponds to the combination of those
    data, with the star indicating the best fit point.\label{fig:app}}
\end{figure}

In Fig.~\ref{fig:app} we show a summary of the $\parenbar{\nu}_\mu \to
\parenbar{\nu}_e$ data in the 3+1 scheme. We observe an allowed region
from MiniBooNE anti-neutrino data that is driven by the event excess
below around 800~MeV and has significant overlap with the parameter
region preferred by LSND. At the 99\%~CL shown in the figure,
MiniBooNE neutrino data give only an upper bound, although we find
closed regions (again driven by the low-energy excess) at lower
confidence levels. This is in qualitative agreement with the results
obtained by the MiniBooNE collaboration, compare Fig.~4
of~\cite{AguilarArevalo:2012va} or Fig.~3
of~\cite{Aguilar-Arevalo:2013ara}.  The different shape of our regions
is due to the oscillations of the background components. Those can be
relatively large in an appearance only fit, since for fixed
$\sin^22\theta_{\mu e}$ we allow $|U_{\mu 4}|$ and $|U_{e 4}|$ to vary
freely, subject to the constraint Eq.~\eqref{eq:sinq2t_me}. We have
checked that when we adopt the same assumptions as the MiniBooNE
collaboration we recover their regions/bounds with good accuracy.

The recent constraint on $\nu_\mu\to\nu_e$ appearance from
ICARUS~\cite{Antonello:2012pq} at long-baseline leads to a bound on
$\sin^22\theta_{\mu e}$ essentially independent of $\Dmq_{41}$ in the
range shown here. It excludes in particular the region of large mixing
and low $\Dmq_{41}$ that is otherwise unconstrained by appearance
experiments.\footnote{Note that this region is also excluded by
  $\nu_e$ and $\nu_\mu$ disappearance searches once
  Eq.~\eqref{eq:sinq2t_me} is used to relate $\sin^2 2\theta_{\mu e}$
  to the effective mixing angles probed by the disappearance
  experiments.} An important background for the $\Dmq_{41}$ driven
$\nu_\mu\to\nu_e$ search in ICARUS are $\nu_e$ appearance events due
to $\Dmq_{31}$ and $\theta_{13}$. Furthermore, as discussed in
section~\ref{sec:parameterisation} and appendix~\ref{app:phases} the
long-baseline appearance probability in the 3+1 scheme depends on one
complex phase.  In deriving the ICARUS bound shown in
Fig.~\ref{fig:app} we fix the parameters $\sin^2 2\theta_{13} = 0.092$
and $\Dmq_{31} = 2.4 \times 10^{-3}~\eVq$ but marginalize over the
relevant complex phase.

As visible in Fig.~\ref{fig:app} there is a consistent overlap region
for all $\parenbar{\nu}_\mu \to \parenbar{\nu}_e$ experiments and we
can perform a combined analysis. The resulting region is shown in red
in Fig.~\ref{fig:app}. The best fit point is at
\fitnumber{$\sin^22\theta_{\mu e} = 0.013$, $\Dmq_{41} = 0.42~\eVq$
  with $\chi^2_\text{min}/\text{dof} = 87.9/(68-2)$~dof (GOF =
  3.7\%)}.  The no-oscillation hypothesis is excluded with respect to
the best fit point with \fitnumber{$\Delta\chi^2 = 47.7$}.  This large
value is mostly driven by LSND.  The relatively low GOF comes mainly
from MiniBooNE neutrino data, as can be seen from
Tab.~\ref{tab:app_chisq}, where we list the individual contribution of
the experiments to the total appearance $\chi^2$. This is also obvious
from Fig.~\ref{fig:mb-spectrum}, showing that at the 3+1 appearance
best fit point (black dotted histogram) the fit to the neutrino
spectrum is not very good, predicting too much excess in the region
$0.6 - 1$~GeV and only partially explaining the excess in the data
below 0.4~GeV.

\begin{table}[t]
  \centering\small
  \fitnumber{
    \begin{tabular}{lr@{/}lr@{/}lr@{/}l}
      \hline\hline
      Experiment & $\chi^2_{3+1}$&dof & $\chi^2_{3+2}$&dof & $\chi^2_{1+3+1}$ & dof \\
      \hline
      LSND            & 11.0&11       &     8.6&11    &  7.5&11 \\
      MiniB $\nu$     & 19.3&11       &    10.6&11    &  9.1&11 \\
      MiniB $\bar\nu$ & 10.7&11       &     9.6&11    & 12.7&11 \\
      E776            & 32.4&24       &    29.2&24    & 31.3&24 \\
      KARMEN          &  9.8&9        &     8.6&9     &  9.0&9  \\
      NOMAD           &  0.0&1        &     0.0&1     &  0.0&1  \\
      ICARUS          &  2.0&1        &     2.3&1     &  1.5&1  \\
      \hline
      Combined        & 87.9&($68-2$) & 72.7&($68-5$) & 74.6&($68-5$) \\
      \hline\hline
  \end{tabular}}
  \mycaption{Individual contributions to the $\chi^2$ at the best fit
    point of the combined appearance data for 3+1, 3+2, and 1+3+1. The
    individual $\chi^2$ values do not add up to the number for the
    combined fit given in the last row because of correlations between
    MiniBooNE neutrino and anti-neutrino data.\label{tab:app_chisq}}
\end{table}

Analysing the same data in the 3+2 scheme we find a best fit point at
\fitnumber{$\Dmq_{41} = 0.57~\eVq$, $\Dmq_{51} = 1.24~\eVq$, with
  $\chi^2_\text{min}/\text{dof} = 72.7/(68-5)$ (GOF = 19\%).}  The GOF
improves considerably with respect to 3+1. We find
\begin{equation}\label{eq:Dchisq-app}
  \fitnumber{
    \chi^2_{3+1,\text{app}} - \chi^2_{3+2,\text{app}} = 15.2 \,.}
\end{equation}
For 3~dof (corresponding to the 3 additional SBL appearance parameters
in 3+2) this implies that appearance data favor 3+2 over 3+1 at the
\fitnumber{99.8\%~CL}.  From Tab.~\ref{tab:app_chisq} we see that
basically all experiments have a reasonable $\chi^2$/dof value (maybe
with the exception of E776, which intrinsically has a somewhat high
$\chi^2$). In particular MiniBooNE neutrino data improve by 8.7 units
compared to 3+1. This is also visible in Fig.~\ref{fig:mb-spectrum},
with the red dotted curve (3+2 appearance best fit) showing a much
better fit than the black dotted one (3+1 appearance best fit), with
$\chi^2 = 24$ for 22~dof for the joint MiniBooNE neutrino and
anti-neutrino data.  The appearance data fit in a 1+3+1 scheme is
similar to the 3+2 case, with a slightly better fit for LSND and
MiniBooNE neutrino, and a slightly worse fit for MiniBooNE
anti-neutrino data, compare Tab.~\ref{tab:app_chisq}. The predicted
MiniBooNE spectra at the 1+3+1 appearance best fit are shown as blue
dotted histograms in Fig.~\ref{fig:mb-spectrum}. We find for 1+3+1
\fitnumber{$\chi^2_\text{min}/\text{dof} = 74.6/(68-5)$ (GOF = 15\%)}
and
\begin{equation}\label{eq:Dchisq-app-1p3p1}
  \fitnumber{
    \chi^2_{3+1,\text{app}} - \chi^2_{1+3+1,\text{app}} = 13.3 \quad (99.6\%\,\text{CL}, \, 3\,\text{dof})}.
\end{equation}

%==============================================================================
\section{Combined analysis of global data}
\label{sec:combi}
%==============================================================================

We now address the question whether the hints for sterile neutrino
oscillations discussed above can be reconciled with each other as well
as with all existing bounds within a common sterile oscillation
framework.  In section~\ref{sec:3+1} we discuss the 3+1 scenario,
whereas in section~\ref{sec:3+2} we investigate the 3+2 and 1+3+1
schemes.

%------------------------------------------------------------------------------
\subsection{3+1 global analysis}
\label{sec:3+1}
%------------------------------------------------------------------------------

\begin{figure}[t]
  \includegraphics[width=0.48\textwidth]{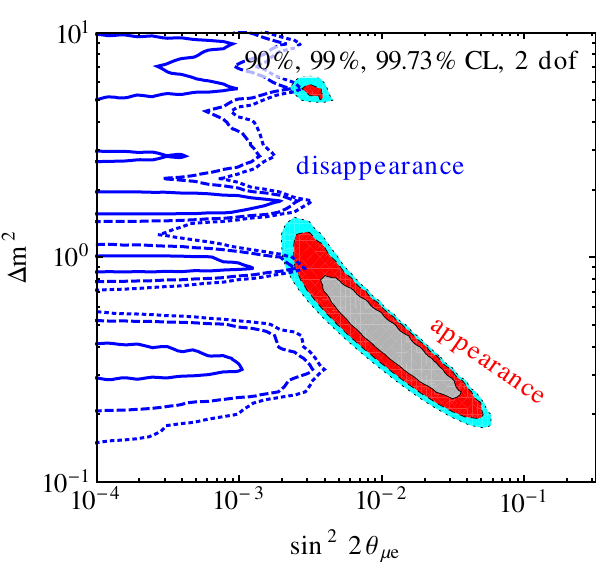}
  \includegraphics[width=0.48\textwidth]{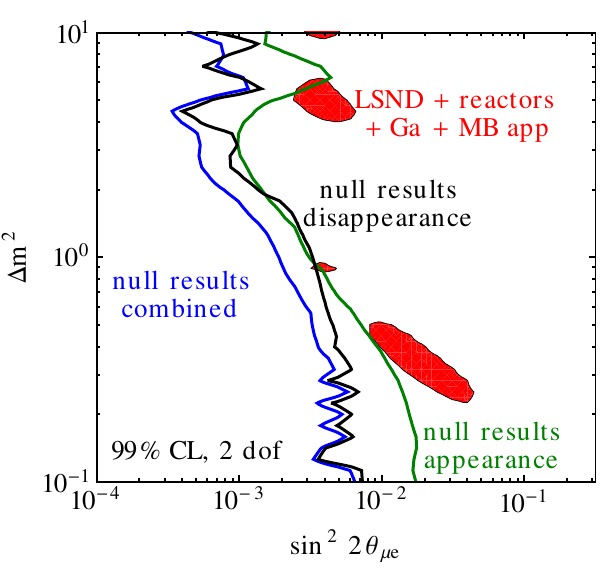}
  \mycaption{Results of the global fit in the $3+1$ scenario, shown as
    exclusion limits and allowed regions for the effective mixing
    angle $\sin^2 2\theta_{\mu e} = 4 |U_{e4}|^2 |U_{\mu 4}|^2$ and
    the mass squared difference $\Dmq_{41}$.  Left: Comparison of the
    parameter region preferred by appearance data (LSND, MiniBooNE
    appearance analysis, NOMAD, KARMEN, ICARUS, E776) to the exclusion
    limit from disappearance data (atmospheric, solar, reactors,
    Gallium, CDHS, MINOS, MiniBooNE disappearance, KARMEN and LSND
    $\nu_e$--$\iso{C}{12}$ scattering). Right: Regions preferred by
    experiments reporting a signal for sterile neutrinos (LSND,
    MiniBooNE, SBL reactors, Gallium) versus the constraints from all
    other data, shown separately for disappearance and appearance
    experiments, as well as their combination.\label{fig:3+1}}
  \end{figure}

In the 3+1 scheme, SBL oscillations are described by effective
2-flavor oscillation probabilities, involving effective mixing angles
for each oscillation channel. The expressions for the effective angles
$\theta_{ee}$, $\theta_{\mu\mu}$, $\theta_{\mu e}$ governing the
$\pbn_e$ disappearance, $\pbn_\mu$ disappearance, and
$\pbn_\mu\to\pbn_e$ appearance probabilities are given in
Eqs.~\eqref{eq:sinq2t_ee}, \eqref{eq:sinq2t_mm}, \eqref{eq:sinq2t_me},
respectively. From those definitions it is obvious that the three
relevant oscillation amplitudes are not independent, since they depend
only on two independent fundamental parameters, namely $|U_{e 4}|$ and
$|U_{\mu 4}|$. Neglecting terms of order $|U_{\alpha 4}|^4$ ($\alpha =
e,\mu$) one finds
\begin{equation}\label{eq:3+1relat}
  \sin^22\theta_{\mu e}
  \approx \frac{1}{4} \sin^22\theta_{ee}\sin^22\theta_{\mu\mu} \,.
\end{equation}
Hence, the appearance amplitude relevant for the LSND/MiniBooNE
signals is quadratically suppressed by the disappearance amplitudes,
which both are constrained to be small. This leads to the well-known
tension between appearance signals and disappearance data in the 3+1
scheme, see \textit{e.g.}~\cite{Bilenky:1996rw,Okada:1996kw} for early
references.

This tension is illustrated for the latest global data in the left
panel of Fig.~\ref{fig:3+1}, where we show the allowed region for all
appearance experiments (the same as the combined region from
Fig.~\ref{fig:app}), compared to the limit from disappearance
experiments in the plane of $\sin^22\theta_{\mu e}$ and $\Delta
m^2_{41}$.  The preferred values of $\Dmq_{41}$ for disappearance data
come from the reactor and Gallium anomalies. The regions for
disappearance data, however, are not closed in this projection in the
parameter space and include $\sin^22\theta_{\mu e} = 4|U_{e 4} U_{\mu
  4}|^2 = 0$, which always can be achieved by letting $U_{\mu 4} \to
0$ because of the non-observation of any positive signal in SBL
$\pbn_\mu$ disappearance. The upper bound on $\sin^22\theta_{\mu e}$
from disappearance emerges essentially as the product of the upper
bounds on $|U_{e 4}|$ and $|U_{\mu 4}|$ from $\pbn_e$ and $\pbn_\mu$
disappearance according to Eq.~\eqref{eq:3+1relat}.  We observe from
the plot the clear tension between those data sets, with only marginal
overlap regions at above 99\%~CL around $\Dmq_{41} \approx 0.9~\eVq$
and at 3$\sigma$ around $\Dmq_{41} \approx 6~\eVq$.

The tension between disappearance and appearance experiments can be
quantified by using the so-called parameter goodness of fit (PG)
test~\cite{Maltoni:2002xd, Maltoni:2003cu}. It is based on the
$\chi^2$ definition
\begin{equation}\begin{split}\label{eq:PG}
    \chi^2_\text{PG}
    &\equiv \chi^2_\text{min,glob} - \chi^2_\text{min,app} - \chi^2_\text{min,dis}
    = \Delta\chi^2_\text{app} + \Delta\chi^2_\text{dis} \,,
    \\[2mm]
    \Delta\chi^2_x &= \chi^2_{x,\text{glob}} - \chi^2_{\text{min},x}
    \quad\text{with}\quad x = \text{app, dis,}
\end{split}\end{equation}
where $\chi^2_\text{min,glob}$ is the $\chi^2$ minimum of the global
data combined, $\chi^2_\text{min,app}$ and $\chi^2_\text{min,dis}$ are
the minima of appearance and disappearance data taken separately, and
$\chi^2_{x,\text{glob}}$ is $\chi^2_x$ evaluated at the best fit point
of the global data.  $\chi^2_\text{PG}$ should be evaluated with the
number of dof corresponding to the number of parameters in common
between appearance and disappearance data (2 in the case of 3+1). From
the numbers given in Tab.~\ref{tab:PG} we observe that the global 3+1
fit leads to $\chi^2_\text{min}$/dof = \fitnumber{712/680} with a
p-value \fitnumber{19\%}, whereas the PG test indicates that
appearance and disappearance data are consistent with each other only
with a p-value of about $10^{-4}$. The strong tension in the fit is
not reflected in the global $\chi^2$ minimum, since there is a large
number of data points not sensitive to the tension, which leads to the
``dilution'' of the GOF value in the global fit,
see~\cite{Maltoni:2003cu} for a discussion. In contrast, the PG test
is designed to test the consistency of different parts of the global
data.

\begin{table}[t]
  \centering\small
  \fitnumber{
    \begin{tabular}{ccccccccc}
      \hline\hline
      & $\chi^2_\text{min}$/dof & GOF
      & $\chi^2_\text{PG}$/dof & PG
      & $\chi^2_\text{app,glob}$ & $\Delta\chi^2_\text{app}$
      & $\chi^2_\text{dis,glob}$ & $\Delta\chi^2_\text{dis}$ \\
      \hline
      3+1    & 712/($689-9$)  & 19\% & 18.0/2 & $1.2\times 10^{-4}$ & 95.8/68 & 7.9 & 616/621 & 10.1 \\
      3+2    & 701/($689-14$) & 23\% & 25.8/4 & $3.4\times 10^{-5}$ & 92.4/68 & 19.7 & 609/621 & 6.1 \\
      1+3+1  & 694/($689-14$) & 30\% & 16.8/4 & $2.1\times 10^{-3}$ & 82.4/68 & 7.8 & 611/621 & 9.0\\
      \hline\hline
  \end{tabular}}
  \mycaption{Global $\chi^2$ minima, GOF values, and parameter
    goodness-of-fit (PG) test~\recite{Maltoni:2003cu} for the
    consistency of appearance versus disappearance experiments in the
    3+1, 3+2, and 1+3+1 schemes.  The corresponding parameter values
    at the global best fit points are given in
    Tab.~\ref{tab:best-fit}. The last four columns give the
    contributions of appearance and disappearance data to
    $\chi^2_\text{PG}$, see Eq.~\eqref{eq:PG}.\label{tab:PG}}
  \end{table}

The conflict between the hints for \eVq-scale oscillations and
null-result data is also illustrated in the right panel of
Fig.~\ref{fig:3+1}. In red we show the parameter regions indicated by
the combined hints for oscillations including SBL reactor, Gallium,
LSND, and MiniBooNE appearance data. Those regions are compared to the
constraint emerging from all other data. We find no overlap region at
99\%~CL. Hence, an explanation of all anomalies within the 3+1 scheme
is in strong tension with constraints from various null-result
experiments.

\begin{table}[t]
  \centering
  \fitnumber{
    \begin{tabular}{cccccccc}
      \hline\hline
      & $\Dmq_{41}$ [\eVq] & $|U_{e4}|$ & $|U_{\mu 4}|$
      & $\Dmq_{51}$ [\eVq] & $|U_{e5}|$ & $|U_{\mu 5}|$
      & $\gamma_{\mu e}$\\
      \hline
      3+1   &  0.93 & 0.15 & 0.17 &        &      &                  \\
      3+2   &  0.47 & 0.13 & 0.15 &   0.87 &  0.14 & 0.13& $-0.15\pi$\\
      1+3+1 & $-0.87$ & 0.15 & 0.13 &   0.47 & 0.13 & 0.17 &  $0.06\pi$\\
      \hline\hline
  \end{tabular}}
  \mycaption{Parameter values at the global best fit points for the
    3+1, 3+2, and 1+3+1 mass schemes. $\gamma_{\mu e}$ is the complex
    phase relevant for SBL appearance experiments as defined in
    Eq.~\eqref{eq:5nu-def}.\label{tab:best-fit}}
  \end{table}

%------------------------------------------------------------------------------
\subsection{3+2 and 1+3+1 global analyses}
\label{sec:3+2}
%------------------------------------------------------------------------------

Now we move to the global analysis within a two-sterile neutrino
scenario in order to investigate whether the additional freedom allows
to mitigate the tension in the fit. We give $\chi^2$ and PG values for
the 3+2 and 1+3+1 schemes in Tab.~\ref{tab:PG} and the corresponding
values of the parameters in Tab.~\ref{tab:best-fit}. We observe from
the PG values that the tension between appearance and disappearance
data remains severe, especially for the 3+2 case, with a PG value
below $10^{-4}$, even less than for 3+1. For 1+3+1 consistency at the
2 per mille level can be achieved.

\begin{figure}[t]
  \centering
  \includegraphics[width=0.6\textwidth]{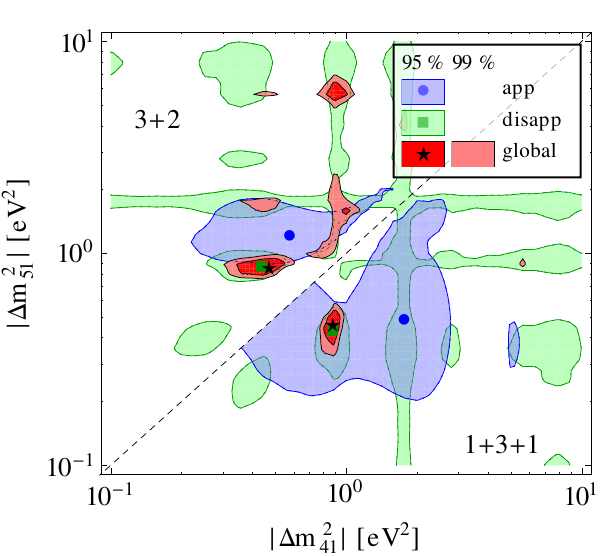}
  \mycaption{Allowed regions in the plane of $|\Dmq_{41}|$ and
    $|\Dmq_{51}|$ in 3+2 (upper-left part) and 1+3+1 (lower-right
    part) mass schemes. We minimize over all mixing angles and
    phases. We show the regions for appearance data (light blue) and
    disappearance data (light green) at 95\%~CL (2~dof), and global
    data (dark and light red) at 95\% and 99\%~CL
    (2~dof).\label{fig:3p2}}
\end{figure}

Let us first discuss the 3+2 fit. We find a modest improvement of the
total $\chi^2$ in the global fit compared to 3+1 by
\begin{equation}\label{eq:Dchisq-glob}
  \fitnumber{
    \chi^2_{3+1,\text{glob}} - \chi^2_{3+2,\text{glob}} =  10.7\,.}
\end{equation}
Evaluated for 4 additional parameters relevant for SBL data in 3+2
compared to 3+1 this corresponds to \fitnumber{96.9\%~CL}.

The origin of the very low parameter goodness of fit can be understood
by looking at the contributions of appearance and disappearance data
to $\chi^2_\text{PG}$. Tab.~\ref{tab:PG} shows that the $\chi^2$ of
appearance data at the global best fit point,
$\chi^2_\text{app,glob}$, changes only by about 3 units between 3+1
and 3+2. However, if appearance data is fitted alone, an improvement
of 15.2 units in $\chi^2$ is obtained when going from 3+1 to 3+2, see
Eq.~\eqref{eq:Dchisq-app}. The fact that appearance data by themselves
are fitted much better in 3+2 than in 3+1 leads to the large value of
$\chi^2_\text{PG} = 25.8$, with a contribution of 19.7 from appearance
data. In other words: the fit to appearance data at the global 3+2
best fit point \fitnumber{($\chi^2_\text{app,glob} = 92.4$/68, p-value
  2.6\%)} is much worse than at the appearance-only 3+2 best fit point
\fitnumber{($\chi^2_\text{min,app}/\text{dof} = 72.7/63$, p-value
  19\%)}. This interpretation is also supported by
Fig.~\ref{fig:mb-spectrum}, showing an equally bad fit to MiniBooNE
neutrino data at the 3+1 and 3+2 global best fit points (black solid
and red solid histograms, respectively).

\begin{figure}[t]
  \includegraphics[width=0.48\textwidth]{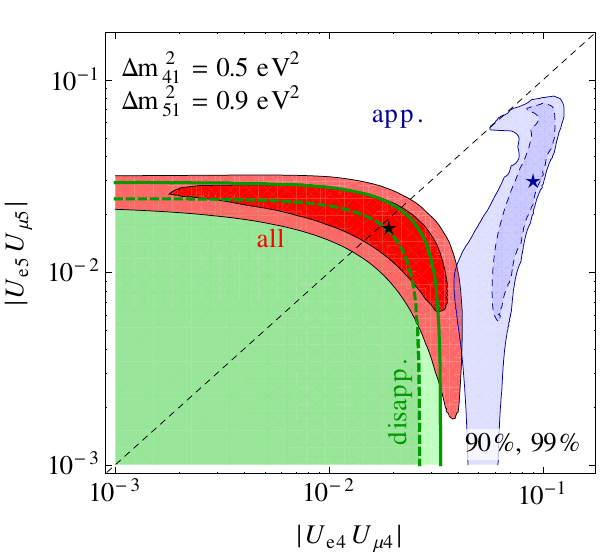}
  \includegraphics[width=0.48\textwidth]{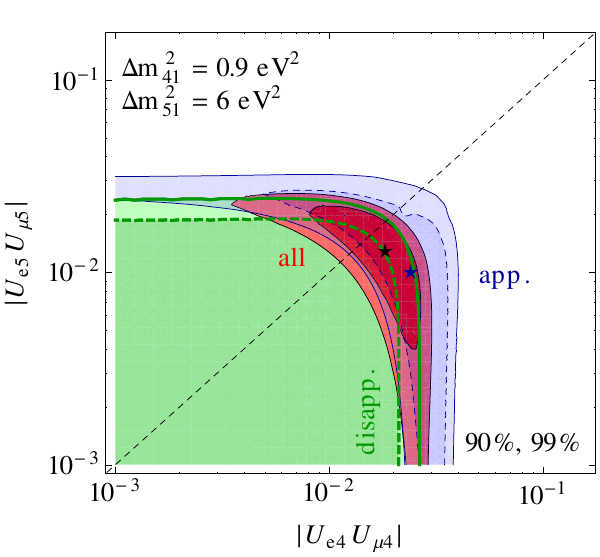}
  \mycaption{Allowed regions for 3+2 in the plane of $|U_{e4} U_{\mu
      4}|$ vs.\ $|U_{e 5} U_{\mu 5}|$ for fixed values of $\Dmq_{41}$
    and $\Dmq_{51}$ at 90\% and 99\%~CL (2~dof). We minimize over all
    undisplayed mixing parameters.  We show the regions for appearance
    data (blue), disappearance data (green), and the global data
    (red).\label{fig:3p2-app-disapp}}
\end{figure}

We further investigate the origin of the tension in the 3+2 fit in
Figs.~\ref{fig:3p2} and~\ref{fig:3p2-app-disapp}. In
Fig.~\ref{fig:3p2} we show the allowed regions in the
multi-dimensional parameter space projected onto the plane of the two
mass-squared differences for appearance and disappearance data
separately, as well as the combined region. The 3+2 global best fit
point happens close to an overlap region of appearance and
disappearance data at 95\%~CL in that plot. However, an overlap in the
projection \emph{does not} imply that the multi-dimensional regions
overlap. In the left panel of Fig.~\ref{fig:3p2-app-disapp} we fix the
mass-squared differences to values close to the global 3+2 best fit
point and show allowed regions in the plane of $|U_{e4}U_{\mu 4}|$ and
$|U_{e 5}U_{\mu 5}|$. These are the 5-neutrino analogs to the
4-neutrino SBL amplitude $\sin 2 \theta_{\mu e}$. Similar as in the
3+1 case we observe a tension between appearance and disappearance
data, with no overlap at 99\%~CL. This explains the small PG
probability at the 3+2 best fit point. The right panel of
Fig.~\ref{fig:3p2-app-disapp} corresponds to the local minimum of the
combined fit visible in Fig.~\ref{fig:3p2} around $\Dmq_{41} =
0.9~\eVq$, $\Dmq_{51} = 6~\eVq$. In this case no tension is visible in
the mixing parameters shown in Fig.~\ref{fig:3p2-app-disapp}, however,
from Fig.~\ref{fig:3p2} we see that those values for the mass-squared
differences are actually not preferred by appearance data, which again
leads to a degraded GOF.  We conclude that the tension between
appearance and disappearance data cannot be resolved in the 3+2
scheme.

\begin{figure}[t]
  \centering
  \includegraphics[width=0.48\textwidth]{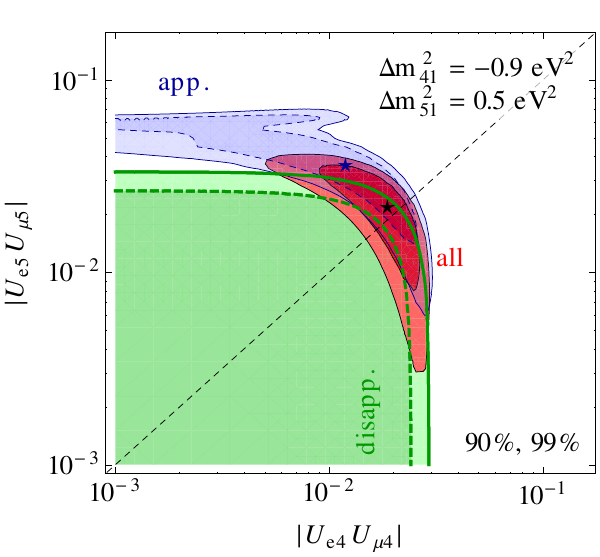}
  \mycaption{Same as Fig.~\ref{fig:3p2-app-disapp} but for the 1+3+1 scheme.\label{fig:1p3p1-app-disapp}}
\end{figure}

For the 1+3+1 ordering of 5-neutrino mass states a somewhat better fit
can be obtained. We find
\begin{equation}\label{eq:Dchisq-1p3p1}
  \fitnumber{
    \chi^2_{3+1,\text{glob}} - \chi^2_{1+3+1,\text{glob}} =  17.8\,,}
\end{equation}
corresponding to disfavoring 3+1 at the \fitnumber{99.9\%~CL} (4~dof)
compared to 1+3+1.  We observe from Tab.~\ref{tab:PG} that at the
1+3+1 global best fit point a much better fit to appearance data is
obtained than at the 3+2 best fit point ($\chi^2_\text{app,glob} =
82.4$ compared to 92.4). As visible from the blue solid histogram in
Fig.~\ref{fig:mb-spectrum} the lack of an event excess in the
MiniBooNE neutrino spectrum around 0.6~GeV is reasonably well
reproduced at the 1+3+1 global best fit point, although the low energy
excess is still under-predicted. The $\chi^2_\text{PG}$ for appearance
versus disappearance for 1+3+1 is even slightly less than for 3+1
(16.8 versus 18.0). Because of the additional parameters relevant for
the evaluation of $\chi^2_\text{PG}$ the p-value 0.2\% is obtained for
1+3+1, about one order of magnitude better than in 3+1.

The projection of the allowed regions on the plane of the mass-squared
differences is shown in the lower-right part of
Fig.~\ref{fig:3p2}. Note that the disappearance regions are to good
accuracy symmetric for 3+2 and 1+3+1. This can be understood from
Eq.~\eqref{eq:SBLdis}, where the difference between 3+2 and 1+3+1
appears only in the last term, which is suppressed by the 4th power of
small matrix elements, compared to the leading terms at 2nd order. We
observe in Fig.~\ref{fig:3p2} that appearance and disappearance
regions for 1+3+1 both overlap with the combined best fit point. In
Fig.~\ref{fig:1p3p1-app-disapp} we show again a section through the
parameter space at fixed values for the mass-squared differences close
to the global best fit point. Although the tension between appearance
and disappearance is still visible (no overlap of the 90\%~CL regions)
the disagreement is clearly less severe than in the 3+2 situation
shown in the left panel of Fig.~\ref{fig:3p2-app-disapp}, and in
Fig.~\ref{fig:1p3p1-app-disapp} we find significant overlap at
99\%~CL, in agreement with the somewhat improved PG p-value.

%==============================================================================
\section{Summary and discussion}
\label{sec:conclusions}
%==============================================================================

We have investigated in detail the status of hints for \eVq-scale
neutrino oscillations, namely the indications for $\pbn_e$
disappearance due to the reactor and Gallium anomalies, and the
indications for $\pbn_\mu\to\pbn_e$ appearance from LSND and
MiniBooNE. Those hints have been analysed in the context of the global
data on neutrino oscillations, including short and long-baseline
accelerator and reactor experiments, as well as atmospheric and solar
neutrinos. Our main findings can be summarized as follows.
\begin{enumerate}
\item For all fits a global $\chi^2_\text{min}/\text{dof} \approx 1$
  is obtained in our analysis, involving 689 data points in total, see
  table~\ref{tab:PG}.

\item However, a joint fit of all anomalies suffers from tension
  between appearance and disappearance data, mainly due to the strong
  constraints from $\pbn_\mu$ disappearance data.

\item The tension in the fit is driven by the LSND and MiniBooNE
  appearance hints, since oscillations in the $\pbn_\mu\to\pbn_e$
  channel inevitably predict also a signal in $\pbn_\mu$
  disappearance, which is not observed at the relevant $L/E$ scale.

\item
In contrast, the reactor and Gallium anomalies are not in direct
conflict with other data, since $\pbn_e$ and $\pbn_\mu$ disappearance at
the \eVq\ scale are controlled by independent parameters.

\item In a 3+1 scheme the compatibility of appearance and
  disappearance data is at the level of $10^{-4}$. The individual
  allowed regions have marginal overlap at about 99\%~CL.

\item We do not find a very significant improvement of the fit in a
  3+2 scheme compared to 3+1. Based on the relative $\chi^2$ minima,
  3+1 is disfavored with respect to 3+2 at 96.9\%~CL. The
  compatibility of appearance and disappearance data in 3+2 is even
  worse than in 3+1, because the fit of appearance data-only is
  significantly better in 3+2 than in 3+1, however, the appearance fit
  at the global best fit point is only marginally improved.

\item We find an improvement of the global fit in the 1+3+1 spectrum
  compared to 3+1, at the 99.9\%~CL. The compatibility of appearance
  and disappearance data is still low in 1+3+1, at the level of 0.2\%.
\end{enumerate}
Hence, in all cases we find significant tension in the fit, with the
marginal exception of the 1+3+1 scheme. At our 1+3+1 best fit point
the minimal value for the sum of all neutrino masses would be $\Sigma
\approx 3\sqrt{|\Dmq_{51}|} + \sqrt{|\Dmq_{41}| + |\Dmq_{51}|} \approx
3.2$~eV, where we took the values given in Tab.~\ref{tab:best-fit} and
assumed that the mass-squared difference with the smaller absolute
value is negative, using the symmetry $4\leftrightarrow 5$ and
$\gamma_{\alpha\beta} \to -\gamma_{\alpha\beta}$ of SBL data, see
Eqs.~\eqref{eq:SBLapp} and~\eqref{eq:SBLdis}. It remains an
interesting question whether such a large value of $\Sigma$ is
consistent with cosmology, see
\textit{e.g.}~\cite{Hamann:2011ge,Giusarma:2011ex,
  GonzalezGarcia:2010un, Joudaki:2012uk, Archidiacono:2013xxa}.

Let us briefly compare our results to two other recent global sterile
neutrino fits, from Refs.~\cite{Conrad:2012qt}
and~\cite{Archidiacono:2013xxa}. We are in good agreement with the
results of~\cite{Conrad:2012qt}. For instance, in Tab.~2
of~\cite{Conrad:2012qt} $\chi^2_\text{PG}$ values for the consistency
of appearance and disappearance data are given, 17.8 for 3+1 and 23.9
for 3+2, which compare well with our numbers from Tab.~\ref{tab:PG},
18.0 and 25.8, respectively. There is some disagreement with the
results of~\cite{Archidiacono:2013xxa}. The corresponding
$\chi^2_\text{PG}$ values reported in their Tab.~I are 6.6 and 11.12,
which lead to significantly better compatibility of appearance and
disappearance data. Comparing Fig.~1 of~\cite{Archidiacono:2013xxa}
with our Fig.~\ref{fig:3+1} (left) we observe that our disappearance
limits are somewhat stronger and our appearance region is at somewhat
larger mixing angles, both effects increasing the tension. Our
appearance region is in good agreement with Fig.~6 (left)
of~\cite{Conrad:2012qt}. There are some differences between our
disappearance region and Fig.~6 (right) of~\cite{Conrad:2012qt},
mainly at high $\Dmq_{41}$.

Irrespective of the hints for $\pbn_e$ disappearance and
$\pbn_\mu\to\pbn_e$ appearance, we have derived constraints on the
mixing of eV-scale states with the $\tau$-neutrino flavor. Those are
dominated by data involving information from neutral-current
interactions, which are solar neutrino data (NC matter effect and SNO
NC data), MINOS long-baseline NC data, and atmospheric neutrino data
(NC matter effect). The global limit is dominated by the latter.

In conclusion, establishing sterile neutrinos at the eV-scale would be
a major discovery of physics beyond the Standard Model. At present a
consistent interpretation of all data indicating the possible presence
of eV-scale neutrino mass states remains difficult. The global fit
suffers from tension between different data sets. An unambiguous
solution to this problem is urgently needed. We are looking forward to
future data on oscillations at the
\eVq\ scale~\cite{Abazajian:2012ys}, as well as new input from
cosmology.

%==============================================================================
\subsection*{Acknowledgments}
%==============================================================================

Numerical results presented in this paper have been obtained on
computing infrastructure provided by Fermi National Accelerator
Laboratory and by Max Planck Institut f\"ur Kernphysik. The authors
would like to thank the MINOS collaboration for their invaluable help
in including their sterile neutrino search in this work. We are
especially grateful to Alexandre Sousa and Mary Bishai for sharing
their Monte Carlo results.  We are grateful to M. Smy for providing
assistance on the simulation of SK4 solar data, and to Bill Louis for
valuable information on the MiniBooNE analysis. Fermilab is operated
by Fermi Research Alliance under contract DE-AC02-07CH11359 with the
US Department of Energy.  P.A.N.M.\ was supported by the
Funda\c{c}\~{a}o de Amparo \`{a} Pesquisa do Estado de S\~{a}o Paulo.
M.M.\ is supported by Spanish MINECO (grants FPA-2009-08958,
FPA-2009-09017, FPA2012-31880, FPA2012-34694, consolider-ingenio 2010
grant CSD-2008-0037 and ``Centro de Excelencia Severo Ochoa'' program
SEV-2012-0249) and by Comunidad Autonoma de Madrid (HEPHACOS project
S2009/ESP-1473). P.A.N.M.\ and M.M.\ acknowledge partial support from
the European Union (FP7 Marie Curie-ITN actions PITN-GA-2009-237920
``UNILHC''). M.M.\ and T.S.\ acknowledge partial support from the
European Union FP7 ITN INVISIBLES (Marie Curie Actions,
PITN-GA-2011-289442).

\appendix

%==============================================================================
\section{Complex phases in sterile neutrino oscillations}
\label{app:phases}
%==============================================================================

In this appendix we discuss in some detail the phases for neutrino
oscillations involving $s$ extra sterile neutrino states. For
definiteness, we will focus on $s=2$; the special case of $s=1$ can be
easily obtained by dropping all terms containing a redundant ``5''
index. Let us order the flavor eigenstates as $(\nu_e,\, \nu_\mu,\,
\nu_\tau,\, \nu_{s_1},\, \nu_{s_2})$ and introduce the following
parametrization for the $n\times n$ mixing matrix, with $n=3+s$:
\begin{equation}\label{eq:Uglob}
  U = V_{35} V_{34}V_{25}V_{24}V_{23}V_{15}V_{14}V_{13}V_{12}
\end{equation}
where $V_{ij}$ represents a complex rotation by an angle $\theta_{ij}$
and a phase $\varphi_{ij}$ in the $ij$ plane.  Note that rotations
involving only sterile states (\textit{i.e.}, $V_{\ell\ell'}$ with
both $\ell, \ell' \ge 4$) are unphysical, and therefore we have
omitted them from Eq.~\eqref{eq:Uglob}. Removing those unphysical
angles, $U$ contains $n(n-1)/2 - s(s-1)/2 = 3(s+1)$ physical angles.

In Eq.~\eqref{eq:Uglob} we have chosen a priori all rotations to be
complex.  We present now a method which allows to remove unphysical
phases from the mixing matrix in a consistent way. First, we note that
a complex rotation can be written as
\begin{equation}\label{eq:rot}
  V_{ij} = D_k O_{ij} D^*_k \,,\qquad k=i \quad\text{or} \quad k=j \,,
\end{equation}
where $O_{ij}$ is a real rotation matrix, $D_k$ is a diagonal matrix
with $(D_k)_{jj} = e^{i\varphi}$ for $j=k$ and $(D_k)_{jj} = 1$ for
$j\neq k$. Depending on whether $k=i$ or $k=j$, the phase in $D_k$ is
either $\pm\varphi_{ij}$. Second, we note that phase matrices $D_k$ at
the very left or right of the matrix $U$ drop out of oscillation
probabilities and are therefore unphysical.\footnote{In this work we
  focus on neutrino oscillations. In cases where lepton-number
  violating processes are relevant, such as neutrino-less double
  beta-decay, more phases will lead to physical consequences and our
  phase counting does not apply. In particular, in such a case the
  phases on the right of the mixing matrix $U$ (these are the
  so-called Majorana phases) cannot be absorbed.} Hence, we have to
represent all matrices $V_{ij}$ in Eq.~\eqref{eq:Uglob} using
Eq.~\eqref{eq:rot}, and then try to commute as many phase matrices to
the left and the right. The matrix $D_k$ commutes with a matrix
$O_{ij}$ if $k \neq i$ and $k\neq j$. Furthermore, if $k = i$ or $k =
j$ we can commute $D_k$ with a complex matrix $V_{ij}$ by re-defining
the phase $\varphi_{ij}$: \textit{e.g.},
$V_{ij}(\theta_{ij},\varphi_{ij}) D_i = D_i
V_{ij}(\theta_{ij},\varphi'_{ij})$. However, we cannot commute $D_k$
with a real matrix $O_{ij}$ if $k = i$ or $k = j$.

This leads to the following rule for removing phases. Let us start by
removing one phase, let's take for instance $\varphi_{12}$, obtaining
a real $V_{12} \to O_{12}$. Then we can no longer use the matrices
$D_1$ and $D_2$ to remove phases, since we cannot commute them with
$O_{12}$ to the very left or right of $U$. But, we can use for
instance $D_3$ to remove one of the remaining phases $\varphi_{i3}$,
and so forth. Hence, we can remove in total $n-1$ phases. Starting
with all $3(s+1)$ physical angles complex, we obtain that there are
$3(s+1) - (n-1) = 2s+1$ physical phases, \textit{i.e.}, 1 phase for no
sterile neutrinos, 3 phases for the 3+1 spectrum, and 5 phases for the
3+2 spectrum. Those remaining phases cannot be associated arbitrarily
to the $V_{ij}$ but only in a way which is consistent with the above
prescription to remove phases. In particular, it is not possible to
make simultaneously three rotation matrices $ij$, $ik$, $kj$ real.
One possible choice is the one given in Eq.~\eqref{eq:rotations}.
Using this recipe to remove phases it is also straightforward to
obtain the physical phases in case of the SBL or LBL approximations
according to Tab.~\ref{tab:counting}.

In the SBL approximation for a 3+2 scheme, only two physical phases
remain. In the parametrization invariant notation from
Eqs.~\eqref{eq:SBLapp} and~\eqref{eq:5nu-def}, they are given by
$\gamma_{\mu e}$ and $\gamma_{\mu\tau}$.  Since the only SBL
appearance experiments we consider are studying the
$\parenbar{\nu}_\mu\to\parenbar{\nu}_e$ oscillation channels only the
phase $\gamma_{\mu e}$ is relevant for our analysis.  In the specific
parametrization from Table~\ref{tab:counting}, the physical phases
have been chosen as $\varphi_{25}$ and $\varphi_{35}$. Since
$\varphi_{35}$ does not appear in the parametrization independent
representation of $\gamma_{\mu e}$ according to Eq.~\eqref{eq:5nu-def}
we can remove it from our SBL analysis without loss of generality.

In the LBL limit, more phases are phenomenologically relevant.  In
particular, Eq.~\eqref{eq:LBLapp} shows that the oscillation
probabilities in the 3+2 case are sensitive to the parametrization
independent phases
\begin{align}\label{eq:phases-3+2}
  \arg(I_{\alpha\beta 43} + I_{\alpha\beta 53}) \,,\qquad
  \arg(I_{\alpha\beta 54}) = \gamma_{\alpha\beta} \,,
\end{align}
with $I_{\alpha\beta ij}$ defined in Eq.~\eqref{eq:5nu-def}.  The
experiments for which the LBL approximation is relevant are ICARUS and
MINOS. ICARUS searches for $\nu_\mu\to\nu_e$ transitions, whereas the
NC data in MINOS are sensitive to the combination $\sum_{\alpha = e,
  \mu, \tau} P_{\nu_\mu \to \nu_\alpha}$. Therefore, for our analyses
the two appearance channels $(\alpha\beta) = (\mu e)$ and $(\mu\tau)$
are relevant, leading, according to Eq.~\eqref{eq:phases-3+2}, to four
independent phases, in agreement with
Tab.~\ref{tab:counting}.\footnote{In deriving Eq.~\eqref{eq:LBLapp} we
  have assumed that $\Dmq_{41},\Dmq_{51},\Dmq_{54}$ are infinite. Note
  that this assumption does not reduce the number of physical phases
  further, since also the general procedure used in
  Tab.~\ref{tab:counting} (assuming only $\Dmq_{21} = 0$) leads to the
  same number of physical phases as Eq.~\eqref{eq:LBLapp}.} The
particular parametrization from the table implies that for the
$\nu_\mu\to\nu_e$ channel only the phases $\varphi_{13}$ and
$\varphi_{25}$ are relevant, whereas the $\nu_\mu\to\nu_\tau$ channel
is also sensitive to $\varphi_{35}$ and $\varphi_{34}$.

From the way we have chosen the complex rotations in
Tab.~\ref{tab:counting} the correct phases in the 3+1 case are
automatically obtained by dropping all rotations including the index
``5'' in the 3+2 mixing matrix. We recover the well-known result that
in the SBL approximation in a 3+1 scenario no complex phase
appears. In the LBL approximation two phases remain, corresponding to
the combinations $\arg(U_{\mu 4}^* U_{e 4} U_{\mu 3} U_{e 3}^*)$ and
$\arg(U_{\mu 4}^* U_{\tau 4} U_{\mu 3} U_{\tau 3}^*)$, which can be
parametrized by using the phases $\varphi_{34}$ and $\varphi_{13}$,
where for the $\nu_\mu\to\nu_e$ channel only $\varphi_{13}$ is
relevant.

Let us comment also on the role of phases in solar and atmospheric
neutrinos. As shown in appendix~\ref{app:solar} solar neutrinos do
depend on one effective complex phase. This is included in our
analysis in full generality however the numerical impact of this phase
dependence is small. It has been shown in~\cite{Maltoni:2007zf}
(appendix~C) that the impact of complex phases on atmospheric
neutrinos is very small and we neglect their effect in the current
analysis.

%==============================================================================
\section{Systematic uncertainties in the reactor analysis}
\label{app:react-errors}
%==============================================================================

The correlation of errors between SBL reactors are quite important in
order to obtain the significance of the reactor anomaly. Here we
describe our error prescription for the SBLR analysis. From the errors
quoted in the original publications we extracted the following
components. First we removed the uncertainty on the neutrino flux
prediction, since we include this uncertainty in a correlated way for
all reactor experiments based on the prescription given
in~\cite{Huber:2011wv} (see below). The remaining error is divided
into uncorrelated errors (including statistical as well as
experimental contributions) as well as correlated errors between some
SBLR measurements. The total uncorrelated error is shown in the last
column of Tab.~\ref{tab:react-data}. Below we give details on our
assumptions on correlations.

%------------------------------------------------------------------------------
%\subsection{Experimental errors for SBL reactor data}
%\label{app:react-errors-exp}
%------------------------------------------------------------------------------

The total error on the measured cross section per fission in Bugey4 is
1.38\%~\cite{Declais:1994ma}. It receives contributions which are
reactor/site specific (1.09\%) as well as detector specific (0.84\%).
Rovno91~\cite{Kuvshinnikov:1990ry} used the same detector as
Bugey4. The errors on the experimental cross section comes from the
reactor and geometry (2.1\%) and the latter from the detector
(1.8\%). So the first one should be uncorrelated whereas the second
one should be correlated with the corresponding one from Bugey4. Hence
we have $\sigma_\text{Bugey4}^\text{uncor} = 1.09\%$,
$\sigma_\text{Bugey4/Rovno91}^\text{cor} = 0.84\%$,
$\sigma_\text{Rovno91}^\text{uncor} = 2.1\%$,
$\sigma_\text{Rovno91/Bugey4}^\text{cor} = 1.8\%$.

The Bugey3 measurement consists of 3 detectors at the distances 15~m,
40~m, 95~m. In Tab.~9 of~\cite{Declais:1994su} systematic errors of
5\% (absolute) and 2\% (relative) are quoted. The uncorrelated errors
given in our Tab.~\ref{tab:react-data} are obtained by adding the
statistical error (Tab.~10 of~\cite{Declais:1994su}) to the 2\%
relative systematic error. For the correlated error we remove the
relative systematic error as well as 2.4\% for the flux prediction and
obtain $\sigma_\text{Bugey3}^\text{cor} = 3.9\%$, which we take fully
correlated between the 3 rate measurements. In cases when we include
the spectral data from Bugey3 we use 2\% (3.9\%) as uncorrelated
(correlated) normalization errors for the three spectra.  Details of
our spectral analysis of Bugey3 can be found in~\cite{Grimus:2001mn}.

In Goesgen the same detector was used at three different distances. In
Tab.~V of~\cite{Zacek:1986cu} the individual and correlated errors are
given. The values for the uncorrelated errors used in our analysis
(see Tab.~\ref{tab:react-data}) are obtained by adding the statistical
and uncorrelated systematic errors in quadrature and expressed in
percentage of the ratio. Then~\cite{Zacek:1986cu} quotes a correlated
error of 6\%, which includes 3\% from the neutrino spectrum, 2\% from
the cross section, 3.8\% from efficiency, 2\% from reactor power, and
a few more $<1$\%. We remove the 3\% neutrino spectrum, as well as the
2\% from cross section (this seems way too large). This gives
$\sigma_\text{Goesgen}^\text{cor} = 4.8\%$.  Part of this error is
supposed to be correlated with ILL, since they used a ``nearly
identical'' detector. Removing the reactor power of 2\% we get
$\sigma_\text{Goesgen/ILL}^\text{cor} = 4.36\%$. In the ILL
paper~\cite{Kwon:1981ua} errors of 3.66\% statistical and 11.5\%
systematical are quoted. The contributions to the systematic error are
given as 6.5\% on the ``intensity of the anti-neutrino energy
spectrum'', 8\% detection efficiency, 1.2\% neutron life time and some
other smaller contributions. In the lack of detailed information we
proceed as follows. We remove 3\% for the flux uncertainties (the same
as in Goesgen) and take 8\% (the detection efficiency) to be
correlated with Goesgen. This gives $\sigma_\text{ILL}^\text{uncor} =
8.52\%$ and $\sigma_\text{ILL/Goesgen}^\text{cor} = 8\%$, where the
uncorrelated error includes also the statistical one. We have checked
that other ``reasonable'' assumptions on the ILL/Goesgen correlation
do not change our results significantly.

From Krasnoyarsk~\cite{Vidyakin:1987ue,Vidyakin:1994ut} there are
three data points based on a single detector, which records events
from 2 ``identical'' reactors.  In~\cite{Vidyakin:1987ue} from 1987,
results at distances of 32.8~m and 92.3~m are reported. The
statistical errors are 3.55\% and 19.8\%, respectively, and the
systematical error are 4.84\% and 4.76\%, respectively, which include
detector effects ($\sim 3\%$), reactor power ($\sim 3\%$) and the
effective distance ($\sim 1\%$). We take systematical errors fully
correlated between those two data points. Then there is a measurement
from 1994~\cite{Vidyakin:1994ut} at 57~m. The errors include detector
uncertainty (3.4\%), reactor power (2.5\%), and statistics
(0.95\%). We assume the detector error to be correlated with the 1987
data points but include the reactor power in the uncorrelated error.

For SRP~\cite{Greenwood:1996pb} measurements at the distances of 18~m
and 24~m are reported from the same detector, which has been moved
between the two positions. The obtained ratios of data over
expectation at the two distances are $98.7\% \pm 0.6\% \text{(stat.)}
\pm 3.7\% \text{(syst.)}$ and $105.5\% \pm 1.0\% \text{(stat.)} \pm
3.7\% \text{(syst.)}$. The uncorrelated systematic error is derived
from the ratio of the two spectra, $1.61 \pm 0.02 \text{(stat)} \pm
0.03 \text{(syst)}$, with an expectation of
1.73~\cite{Greenwood:1996pb}. Hence $1.86\% = 0.03/1.61$ is an
uncorrelated systematic error. Then we remove the 2.5\% from the
neutrino spectrum from the systematical error and obtain
$\sigma_\text{SRP1}^\text{uncor} = 1.95\%$,
$\sigma_\text{SRP2}^\text{uncor} = 2.11\%$, and
$\sigma_\text{SRP}^\text{cor} = 2.0\%$. With this assumption on the
uncorrelated errors the two data points are consistent at about
$2.4\sigma$.

Rovno88~\cite{Afonin:1988gx} reports 5 measurements with two different
detectors: 1I, 2I, 1S, 2S, 3S, where the ``I'' experiments use an
integral neutron detector, whereas the ``S'' experiments use a
scintillation detector measuring the positron spectrum. In Tab.~III
of~\cite{Afonin:1988gx} for each measurement two systematical errors
are given, 2.2\% for ``the uncertainty in the measured reactor power
and the geometric uncertainty'', and a second uncertainty due to
``errors in the detector characteristics and fluctuations''. From
Tab.~II one finds that statistical errors are negligible. In the
absence of detailed information we assume the 2.2\% uncertainty fully
correlated among all experiments. From the second error we assume that
half of it is uncorrelated and the other half is correlated among
detectors of the same type. We have checked that our results do not
depend significantly on those assumptions.

%------------------------------------------------------------------------------
%\subsection{Uncertainties on the neutrino fluxes}
%\label{app:react-errors-flux}
%------------------------------------------------------------------------------

\bigskip

Finally let us comment on the uncertainty on the neutrino flux
predictions.  As mentioned above this uncertainty has been removed
from the SBLR experimental errors since they are treated in a
correlated way for all reactor experiments. For the uncertainties of
the fluxes from $^{235}$U, $^{239}$Pu, $^{241}$Pu we use the
information from tables provided in~\cite{Huber:2011wv}. The
uncertainty is provided as uncorrelated error in each bin of neutrino
energy as well as fully correlated (between energy bins as well as the
three isotopes) errors. For the uncorrelated errors we proceed as
follows. We perform a fit of a polynomial of 2nd order to the numbers
given in~\cite{Huber:2011wv}. Then those coefficients are used as
pulls in the $\chi^2$ analysis constrained by the covariance matrix
obtained from the polynomial fits. This allows us to take into account
the fact that the bin-to-bin uncorrelated errors of the neutrino
spectrum will lead to correlated effects in the observed positron
spectra. Since the uncorrelated flux errors are sub-leading compared
to the correlated ones the parametrization with a 2nd order polynomial
is sufficiently accurate.  To include the correlated errors we
follow~\cite{Huber:2011wv}: the various contributions to this error in
each neutrino energy bin are symmetrized and added in quadrature. Then
we obtain an energy dependent fully correlated error for the spectra
from $^{235}$U, $^{239}$Pu, $^{241}$Pu which is included as one common
pull parameter in the global reactor $\chi^2$. For the neutrinos from
$^{238}$U we use the flux from~\cite{Mueller:2011nm} and include a
global normalization error on the $^{238}$U induced events of
8.15\%~\cite{Mention:2011rk}.

%==============================================================================
\section{Solar neutrino analysis}
\label{app:solar}
%==============================================================================

In the analysis of solar neutrino experiments we include the total
rates from the radio chemical experiments
Chlorine~\cite{Cleveland:1998nv}, GALLEX/GNO~\cite{Kaether:2010ag} and
SAGE~\cite{Abdurashitov:2009tn}.  Regarding real-time experiments, we
include the electron scattering energy-zenith angle spectrum data from
all the Super-Kamiokande phases I--IV~\cite{Hosaka:2005um,
  Cravens:2008aa, Abe:2010hy, SKnu2012} and the data from the three
phases of SNO~\cite{Aharmim:2007nv, Aharmim:2005gt, Aharmim:2008kc}.
We also include the main set of the 740.7~days of Borexino
data~\cite{Bellini:2011rx} as well as their high-energy spectrum from
246~live days~\cite{Bellini:2008mr}. In total the solar neutrino data
used in our analysis consists of 261 data points.

Let us now focus on the probabilities relevant for the analysis of
solar neutrino experiments. We will assume that only the first two
mass eigenstates are dynamical, while the others are taken to be
infinite. Since physical quantities have to be independent of the
parameterization of the mixing matrix, we will use the freedom in
choosing a parameterization that makes analytical expressions
particularly simple. We start from the Hamiltonian in the flavor
basis:
\begin{equation}\label{eq:Hflavor}
  H = U \Delta U^\dagger + V \,,
\end{equation}
where $\Delta = \diag(0,\, \Dmq_{21},\, \Dmq_{31},\, \dots) / 2E$ and
$V = \sqrt{2} \, G_F \diag(2 N_e,\, 0,\, 0,\, N_n,\, \dots)/2$. It is
convenient to write $U = \tilde{U} U_{12}$, where $U_{12}$ is a
complex rotation by an angle $\theta_{12}$ and a phase $\delta_{12}$
which we will define later.\footnote{We define $\theta_{12}$ in such a
  way that the corresponding rotation is the rightmost rotation in
  $U$, but we do not assume anything about all other rotations. If the
  standard parameterization is used for them, then $\theta_{12}$ has
  the usual interpretation.} Then we can write:
\begin{equation}
  H = \tilde{U} \tilde{H} \tilde{U}^\dagger
  \qquad\text{with}\qquad
  \tilde{H} = U_{12} \Delta U_{12}^\dagger
  + \tilde{U}^\dagger V \tilde{U} \,.
\end{equation}
In order to further simplify the analysis, let us now assume that all
the mass-squared differences involving the ``heavy'' states $\nu_h$
with $h \ge 3$ can be considered as infinite: $\Dmq_{hl} \to \infty$
and $\Dmq_{hh'} \to \infty$ for any $l=1,2$ and $h,h' \ge 3$.  In
leading order, the matrix $\tilde{H}$ takes the effective
block-diagonal form:
\begin{equation}
  \tilde{H} \approx
  \begin{pmatrix}
    H^{(2)} & \boldsymbol{0} \\
    \boldsymbol{0} & \Delta^{(s)}
  \end{pmatrix}
\end{equation}
where $H^{(2)}$ is the $2\times 2$ sub-matrix of $\tilde{H}$
corresponding to the first and second neutrino states, and
$\Delta^{(s)} = \diag(\Dmq_{31},\, \Dmq_{41},\, \dots)/ 2E$ is a
diagonal $(s+1)\times (s+1)$ matrix (the matter terms in this block
are negligible in the limit of very large $\Dmq_{hh'}$).
Consequently, the evolution matrix is:
\begin{equation}
  \tilde{S} \approx
  \begin{pmatrix}
    S^{(2)} & \boldsymbol{0} \\
    \boldsymbol{0} & e^{-i \Delta^{(s)} L}
  \end{pmatrix}
  \quad\text{with}\quad
  S^{(2)} = \Evol\big[H^{(2)}\big]
  \quad\text{and}\quad
  S = \tilde{U} \tilde{S} \tilde{U}^\dagger \,.
\end{equation}
We are interested only in the elements $S_{\alpha e}$. It is
convenient to define $\theta_{12}$ in such a way that $\tilde{U}_{e2}
= 0$. Taking into account the block-diagonal form of $\tilde{S}$, we
obtain:
\begin{equation}
  S_{\alpha e} = \tilde{U}_{e1}^\star
  \left( \tilde{U}_{\alpha 1} S_{11}^{(2)}
  + \tilde{U}_{\alpha 2} S_{21}^{(2)} \right)
  + \sum_{h\ge 3} \tilde{U}_{\alpha h} \tilde{U}_{eh}^\star
  e^{-i\Delta_{hh}L} \,.
\end{equation}
The expressions for the probabilities, $P_{\alpha e} = |S_{\alpha
  e}|^2$, are straightforward:
\begin{equation}
  \begin{split}
    P_{\alpha e} &= |\tilde{U}_{e1}|^2
    \left[ |\tilde{U}_{\alpha 1}|^2 |S_{11}^{(2)}|^2
      + |\tilde{U}_{\alpha 2}|^2 |S_{21}^{(2)}|^2
      + 2 \Re\big( \tilde{U}_{\alpha 1} \tilde{U}_{\alpha 2}^\star
      S_{11}^{(2)} S_{21}^{(2)\star} \big) \right]
    + \sum_{h\ge 3} |\tilde{U}_{\alpha h}|^2 |\tilde{U}_{eh}|^2
    \\
    &= |\tilde{U}_{e1}|^2 \left[ \big(
      |\tilde{U}_{\alpha 2}|^2 - |\tilde{U}_{\alpha 1}|^2 \big) |S_{21}^{(2)}|^2
      + 2 \Re\big( \tilde{U}_{\alpha 1} \tilde{U}_{\alpha 2}^\star
      S_{11}^{(2)} S_{21}^{(2)\star} \big) \right]
    + \sumAll |\tilde{U}_{\alpha i}|^2 |\tilde{U}_{ei}|^2 \,.
  \end{split}
\end{equation}
Here we have used the fact that the terms containing a factor
$e^{-i\Delta_{hh}L}$ oscillate very fast, and therefore vanish once
the finite energy resolution of the detector is taken into
account. For solar neutrino experiments we only need $P_{ee}$ and
$P_{ae} \equiv P_{ee} + P_{\mu e} + P_{\tau e}$. It is therefore
convenient to define $\delta_{12}$ in such a way that $\sumSte
\tilde{U}_{\sigma 1} \tilde{U}_{\sigma 2}^\star$ is a real number. Let
us also define:
\begin{equation}
  \eta_e \equiv |\tilde{U}_{e1}|^2,
  \qquad
  \xi_D \equiv \sumSte \big(
  |\tilde{U}_{\sigma 2}|^2 - |\tilde{U}_{\sigma 1}|^2 \big),
  \qquad
  \xi_N \equiv 2 \sumSte
  \tilde{U}_{\sigma 1} \tilde{U}_{\sigma 2}^\star \,.
\end{equation}
Using unitarity relations, $\sum_\alpha |\tilde{U}_{\alpha i}|^2 = 1$
and $\sum_\alpha \tilde{U}_{\alpha 1} \tilde{U}_{\alpha 2}^\star = 0$,
we obtain:
\begin{align}
  P_{ee} = \tilde{C}_e - \eta_e^2 P_\text{osc}^{(2)} \,,
  \qquad
  P_{ae} = \tilde{C}_a - \eta_e \left(
  \xi_D P_\text{osc}^{(2)} + \xi_N P_\text{int}^{(2)} \right)
\end{align}
where
\begin{equation}
  \begin{gathered}
    P_\text{osc}^{(2)} \equiv |S_{21}^{(2)}|^2 \,,
    \qquad
    P_\text{int}^{(2)} \equiv \Re\big(
    S_{11}^{(2)} S_{21}^{(2)\star} \big) \,,
    \\[1mm]
    \tilde{C}_e \equiv \sumAll |\tilde{U}_{ei}|^4 \,,
    \qquad
    \tilde{C}_a \equiv 1 - \sumAll \sumSte
    |\tilde{U}_{ei}|^2 |\tilde{U}_{\sigma i}|^2 \,.
  \end{gathered}
\end{equation}
In the above expressions $P_\text{osc}^{(2)}$ and $P_\text{int}^{(2)}$
are effective terms derived from the Hamiltonian $H^{(2)}$, which has
the form:
\begin{equation}
  \begin{split}
    H^{(2)} &= H_\text{vac}^{(2)}
    + \sqrt{2} \, G_F \, N_e
    \begin{pmatrix}
      |\tilde{U}_{e1}|^2
      & \tilde{U}_{e1}^\star \tilde{U}_{e2}
      \\
      \tilde{U}_{e1} \tilde{U}_{e2}^\star
      & |\tilde{U}_{e2}|^2
    \end{pmatrix}
    + \sqrt{2} \, G_F \, \frac{N_n}{2} \, \sumSte
    \begin{pmatrix}
      |\tilde{U}_{\sigma 1}|^2
      & \tilde{U}_{\sigma 1}^\star \tilde{U}_{\sigma 2}
      \\
      \tilde{U}_{\sigma 1} \tilde{U}_{\sigma 2}^\star
      & |\tilde{U}_{\sigma 2}|^2
    \end{pmatrix}
    \\
    &= H_\text{vac}^{(2)}
    + \sqrt{2} \, G_F \, N_e \, \eta_e
    \begin{pmatrix}
      1 & 0
      \\
      0 & 0
    \end{pmatrix}
    + \sqrt{2} \, G_F \, \frac{N_n}{4}
    \begin{pmatrix}
      -\xi_D & ~\xi_N
      \\
      \hphantom{-}\xi_N & ~\xi_D
    \end{pmatrix}
  \end{split}
\end{equation}
with the vacuum term including the phase $\delta_{12}$:
\begin{equation}
  H_\text{vac}^{(2)} =
  \frac{\Dmq_{21}}{4E}
  \begin{pmatrix}
    -\cos2\theta_{12} \, \hphantom{e^{-i\delta_{12}}}
    & ~\sin2\theta_{12} \, e^{i\delta_{12}}
    \\
    \hphantom{-}\sin2\theta_{12} \, e^{-i\delta_{12}}
    & ~\cos2\theta_{12} \, \hphantom{e^{i\delta_{12}}}
  \end{pmatrix} .
\end{equation}
These equations are valid for any number of extra sterile
states. Hence, in the most general case solar neutrinos depend on six
real mixing parameters,
\begin{equation}\label{eq:mixing_sol}
  \theta_{12},\, \eta_e,\, \xi_D,\, \xi_N, \, \tilde{C}_e, \, \tilde{C}_a \,,
\end{equation}
one complex phase ($\delta_{12}$), and
$\Dmq_{21}$. Let us now further define $\eta_s \equiv
\sqrt{\xi_D^2 + \xi_N^2}$ and discuss some special limits:
\begin{itemize}
\item $\eta_e = 1$: this implies $\tilde{U}_{e1} = 1$, hence $U_{ei} =
  0$ for $i \ge 3$. By unitarity $\sum_\alpha |\tilde{U}_{\alpha 1}|^2
  = 1$, hence $\tilde{U}_{\alpha 1} = 0$ for $\alpha \ne
  e$. Therefore, $\tilde{C}_e = \tilde{C}_a = 1$ and $\xi_N = 0$; the
  probabilities reduce to the well-known expressions $P_{ee} = 1 -
  P_\text{osc}^{(2)}$ and $P_{ae} = 1 - \eta_s P_\text{osc}^{(2)}$
  with $\eta_s = \xi_D$;

\item $\eta_s = 0$: in this case $\xi_D = \xi_N = 0$, sterile
  neutrinos do not participate to 1-2 oscillations and $P_{ae}$ is a
  constant. If we further assume that no sterile neutrino exists then
  $\tilde{C}_e = \eta_e^2 + (1 - \eta_e)^2$ and $\tilde{C}_a = 1$, so
  that $P_{ae} = 1$ and $P_{ee}$ reduces to the well-known
  three-neutrino formula;

\item $s = 1$: for one sterile neutrino $\eta_s$ reduces to $\eta_s =
  |\tilde{U}_{s1}|^2 + |\tilde{U}_{s2}|^2$.
\end{itemize}

To perform the analysis we map the parameters which are used for the
analysis into the effective parameters for solar neutrinos. In the
case of 3+1 the number of real mixing parameters is actually the same
as the number of mixing angles in the most general parametrization of
the $4\times 4$ mixing matrix: the six parameters in
Eq.~\eqref{eq:mixing_sol} are a function of the six angles
$\theta_{12}$, $\theta_{13}$, $\theta_{14}$, $\theta_{23}$,
$\theta_{24}$, $\theta_{34}$.  The dependence of solar neutrinos on
$\theta_{13},\theta_{14}$ is shown in Fig.~\ref{fig:th13-th14} and the
one on $\theta_{24},\theta_{34}$ follows from
Fig.~\ref{fig:th24th34}. The dependence on $\theta_{23}$ is important
for the NC matter effect and SNO NC data. For $s > 1$ the number of
effective mixing parameters of solar data is less than the number of
angles in the general mixing matrix. The phase $\delta_{12}$ is a
complicated function of complex phases and angles. We have verified
numerically for the 3+1 case that if all angles are non-zero (and
$\theta_{24},\theta_{34}$ relatively large) the $\chi^2$ from solar
data varies by about 1 to 2 units as a function of the phase. Once all
relevant constraints on the mixing angles are imposed the effect of
the phase on solar data is negligible. See also~\cite{Long:2013hwa}
for a discussion of complex phases in solar neutrinos in the context
of sterile neutrinos.

%==============================================================================
\section{Atmospheric neutrino analysis}
\label{app:atm}
%==============================================================================

The analysis of atmospheric data follows closely the one presented in
Refs.~\cite{GonzalezGarcia:2010er, GonzalezGarcia:2011my} and includes
the Super-Kamiokande results from phases I, II and
III~\cite{Wendell:2010md} (80 data points in total). Technical details
on our $\chi^2$ fit can be found in the appendix
of~\cite{GonzalezGarcia:2007ib}.

In order to derive suitable expressions for the relevant
probabilities, we can follow the approach presented in
App.~\ref{app:solar} for solar data. The Hamiltonian in the flavor
basis is given by Eq.~\eqref{eq:Hflavor}, which in the mass basis
becomes:
\begin{equation}
  H = U \hat{H} U^\dagger
  \qquad\text{with}\qquad
  \hat{H} = \Delta
  + U^\dagger V U \,.
\end{equation}
As before, we can simplify the analysis by assuming that all the
mass-squared differences involving the ``heavy'' states $\nu_h$ with
$h \ge 4$ can be considered as infinite: $\Dmq_{h i} \to \infty$ and
$\Dmq_{hh'} \to \infty$ for any $i=1,2,3$ and $h,h' \ge 4$.  In
leading order, the matrix $\hat{H}$ takes the effective block-diagonal
form:
\begin{equation}
  \hat{H} \approx
  \begin{pmatrix}
    \hat{H}^{(3)} & \boldsymbol{0} \\
    \boldsymbol{0} & \Delta^{(s)}
  \end{pmatrix}
\end{equation}
where $\hat{H}^{(3)}$ is the $3\times 3$ sub-matrix of $\hat{H}$
corresponding to the first, second and third neutrino states, and
$\Delta^{(s)} = \diag(\Dmq_{41},\, \Dmq_{51},\, \dots)/ 2E_\nu$ is a
diagonal $s \times s$ matrix (the matter terms in this block are
negligible in the limit of very large $\Dmq_{hh'}$).  We are
interested only in the probabilities $P_{\alpha\beta}$ with $\alpha,
\beta \in \{ e, \mu, \tau \}$. Taking into account the block-diagonal
form of $\hat{H}$, we obtain:
\begin{equation}
  P_{\alpha\beta} = \big| A_{\alpha\beta} \big|^2 + K_{\alpha\beta} \,,
  \qquad
  A = U^{(3)} \hat{S}^{(3)} U^{(3)\dagger} \,,
  \qquad
  K_{\alpha\beta} \equiv \sum_{h\ge 4} |U_{\alpha h}|^2 |U_{\beta h}|^2
\end{equation}
where $\hat{S}^{(3)} = \Evol\big[ \hat{H}^{(3)} \big]$ and $U^{(3)}$
is the $3 \times 3$ sub-matrix corresponding to the first three lines
and columns of $U$, and as such it is \emph{not} a unitary matrix.

Let us now focus on the three-neutrino system described by
$\hat{H}^{(3)} = \Delta^{(3)} + \hat{V}^{(3)}$, with $\hat{V}^{(3)} =
[U^\dagger V U]^{(3)}$. The matter term $\hat{V}^{(3)}$ contains both
the ``standard'' contribution from $\nu_e$ charged-current
interactions, and a ``non-standard'' part induced by the absence of
neutral-current interactions for sterile neutrinos.  In practical
terms, this is just the same as the problem of neutrino propagation in
the presence of non-standard neutrino-matter interactions (NSI)
described in Ref.~\cite{GonzalezGarcia:2011my}. Following the approach
discussed there, we make a number of simplifying assumptions:
\begin{itemize}
\item we assume that the neutron-to-electron density ratio, $R_{ne}$,
  is constant all over the Earth. We set $R_{ne} = 1.051$ as inferred
  from the PREM model~\cite{Dziewonski:1981xy};

\item we set $\Dmq_{21} = 0$, thus forcing the vacuum term
  $\Delta^{(3)}$ to have two degenerate eigenvalues;

\item we impose that the matter term $\hat{V}^{(3)}$ also has two
  degenerate eigenvalues.
\end{itemize}
The first two assumptions are very well known hand have been discussed
in detail in the literature. For example, in Sec.~5.2 of
Ref.~\cite{GonzalezGarcia:2007ib} it was noted that the different
chemical composition of the Earth mantle and core has very little
impact on NSI results. The last approximation is adopted here for
purely practical reasons, since (together with the other two) it
allows to greatly simplify the calculation of the neutrino
evolution~\cite{Blennow:2008eb, GonzalezGarcia:2011my}. Unfortunately,
in the present context the matter term $\hat{V}^{(3)}$ cannot be fixed
\textit{a priori}, but it arises as an effective quantity determined
by the angles and phases of the mixing matrix $U$. Finding the points
in the general parameter space for which $\hat{V}^{(3)}$ has two
degenerate eigenvalues (the only points for which our numerical
analysis is technically feasible) is not an easy task. To solve this
problem, we have considered here two alternative cases, both based on
physically motivated scenarios:
\begin{enumerate}
\item[(\emph{a})] decouple the electron flavor from the evolution and
  include the NC matter effect;

\item[(\emph{b})] allow the electron flavor to participate in
  oscillations but neglect the NC matter effect.
\end{enumerate}
Option (\emph{a}) requires to set $U_{ei}=0$ for $i \ge 3$ (in
addition to $\Dmq_{21} = 0$), whereas option (\emph{b}) is equivalent
to setting $R_{ne} = 0$ (``hydrogen-Earth'' model). Both
approximations have been previously discussed in App.~C of
Ref.~\cite{Maltoni:2007zf}.  Although we know that none of these
options corresponds to Nature, we can make a sensible choice of when
it is safe to use each of them. It turns out that for constraining the
mixing of the $\nu_\mu$ with eV-scale neutrinos the NC matter effect
plays no role at all (see discussion in App.~C2
of~\cite{Maltoni:2007zf}), whereas the participation of the electron
flavor may have some impact.\footnote{The reason is that the main
  information comes from effects on the overall normalization of
  $\mu$-like events, and in order to fix this normalization the
  $e$-like sample plays a role in constraining uncertainties in the
  neutrino fluxes.}  Therefore, whenever we are mainly interested in
constraining $|U_{\mu h}|$ ($h \ge 4$), as in the case of the global
analysis combined with SBL data, we adopt assumption (\emph{b}).  This
is important since non-zero $|U_{e i}|$ leads to slightly relaxed
constraints on $|U_{\mu 4}|$, although the effect is small once
external constraints on $|U_{e i}|$ are taken into account. With
respect to our former analysis presented in
Ref.~\cite{Maltoni:2007zf}, the explicit NSI formalism adopted here is
more general since it allows to fully include $|U_{e i}|$-related
effects without further approximations.

On the other hand, when exploring the sensitivity to the fraction of
sterile neutrinos participating in atmospheric and long-baseline
neutrino oscillations, the contribution of neutral-current
neutrino-matter interactions is essential. Indeed, under approximation
(\emph{b}) no limit on $|U_{\tau i}|$ ($i \ge 4$) would be
obtained. Therefore, when exploring constraints on $|U_{\tau i}|$ we
adopt assumption (\emph{a}) above. This is relevant for
Figs.~\ref{fig:th24dm41} (right) and~\ref{fig:th24th34}.

%==============================================================================
\section{Technical details on the simulation of SBL and LBL experiments}
\label{app:exp-details}
%==============================================================================

Here we provide technical details on the simulation of some of the
experiments included in our fit. All the simulations described in this
appendix make use of the GLoBES software package~\cite{Huber:2004ka,
  Huber:2007ji}.

%==============================================================================
\subsection{KARMEN/LSND $^{12}$C $\nu_e$ disappearance analysis}
\label{app:carbon}
%==============================================================================

Both LSND and KARMEN have measured the reaction $\nu_e +
{}^{12}\text{C} \to \text{e}^- + {}^{12}\text{N}$, where the
${}^{12}$N decays back to ${}^{12}\text{C} + \text{e}^+ + \nu_e$ with
a lifetime of 15.9~ms. By detecting the electron from the first
reaction, which has a $Q$-value of 17.33~MeV, one can infer the
neutrino energy. Here we describe how we use these data to constrain
$\nu_e$ disappearance~\cite{Reichenbacher:2005nc, Conrad:2011ce}.

For KARMEN, we use the information from the
thesis~\cite{Reichenbacher:2005nc}, which uses more exposure than the
original publication~\cite{Armbruster:1998uk}. The number of expected
events can be calculated by multiplying the $^{12}$C cross section
(Fukugita et al.~\cite{Fukugita:1988hg}: $9.2 \pm 1.1 \times
10^{-42}~\text{cm}^2$), the number of target nuclei ($2.54 \times
10^{30}$), the absolute neutrino flux ($5.23\times 10^{21}$), the
efficiency (27.2\%, flat in energy), and the inverse effective scaled
area ($1/[4 \pi (17.72 \text{m})^2]$).
846 neutrino candidates are observed, with an expected background of
$13.9\pm0.7$, which are mainly accidentals and cosmic induced.  The
systematic errors are dominated by a 6.7\% uncertainty in the absolute
neutrino flux and 3\% in the Monte Carlo efficiency. The total
systematic error is 7.5\% plus a 12\% cross section error. We take the
latter to be correlated between the KARMEN and LSND $\nu_e$-carbon
analyses.
In fig.~3.2 (upper panel) of~\cite{Reichenbacher:2005nc}, the data are
shown as the energy distribution of the prompt $e^-$ spectrum in 26
bins where the visible prompt electron energy is within the range
$10~\text{MeV} < E_{e} < 36~\text{MeV}$. Modulo energy reconstruction
the neutrino and electron energies would be related by the $Q$-value
as $E_\nu = E_e + Q$.  For the simulation we assume $30~\text{MeV} <
E_{\nu} < 56~\text{MeV}$. To properly fit the data, we assume
$\sigma_e = 25\%/\sqrt{E~\text{(MeV)}}$ for the energy
resolution. With the 26 data points we obtain a two-neutrino best fit
with $\chi^2_\text{min} / \text{dof} = 30/24$.

For LSND~\cite{Auerbach:2001hz} we compute the expected number of
events by multiplying the $^{12}$C cross section ($9.2 \times
10^{-42}~\text{cm}^2$~\cite{Fukugita:1988hg}), the number of target
nuclei ($3.34 \times 10^{30}$), the neutrino flux at the detector
($10.58\times 10^{13}~\text{cm}^{-2}$), and the efficiency (23.2\%,
flat in energy).
733 neutrino candidates are observed, with a negligible expected
background.  The systematic error is dominated by a 7\% uncertainty in
the neutrino flux and a 6\% uncertainty in the effective fiducial
volume. The total systematic error, not including the theoretical
cross section error, is 9.9\%. The 12\% cross section error is
correlated between the LSND and KARMEN $\nu_e$-${}^{12}\text{C}$
analyses. In fig.~6 of Ref.~\cite{Auerbach:2001hz}, the data are shown
as the energy distribution of the prompt $e^-$ spectrum, where the
visible prompt electron energy is within the range $18~\text{MeV} <
E_{e} < 42~\text{MeV}$ and divided into 12 bins of width 2~MeV. In
terms of the neutrino energy $E_\nu$, this energy range corresponds to
$35.3~\text{MeV} < E_{\nu} < 59.3~\text{MeV}$.  In our analysis, we
combine the 12 energy bins into only 6 bins.  To properly fit the
data, we assume $\sigma_e = 2.7$~MeV for the energy resolution. With
the 6 data points we obtain a two-neutrino best fit with
$\chi^2_\text{min} / \text{dof} = 3.81/4$.

For the combined KARMEN+LSND $\nu_e$-carbon fit, we have 32 bins and
we find a two-neutrino best fit point with $\chi^2_\text{min} /
\text{dof} = 34.17/30$.

%------------------------------------------------------------------------------
\subsection{E776}
\label{app:E776}
%------------------------------------------------------------------------------

The pion beam experiment E776 at Brookhaven~\cite{Borodovsky:1992pn}
employed a 230~ton calorimeter detector located at approximately 1~km
from the end of the 50~m long pion decay pipe. The $\parenbar\nu_e$
energy of order GeV was measured with an energy resolution of
20\%/$\sqrt{E\,\text{[GeV]}}$.  E776 used $\parenbar\nu_\mu$
disappearance data to obtain the overall normalization of the neutrino
flux. In our fit, we do not explicitly include $\parenbar\nu_\mu$
data, but instead use the normalization as an input.  The main
backgrounds in E776 came from intrinsic $\parenbar\nu_e$ contamination
in the beam and from $\pi^0$'s produced in neutral current
interactions and misidentified as electrons.  The systematic errors
were 11\% for the intrinsic background and 27\% (39\%) for the $\pi^0$
background in neutrino (anti-neutrino) mode.  In 1986, E776 collected
$1.43 \times 10^{19}$ ($1.55 \times 10^{19}$) protons on target, and a
total of 136 (56) $\nu_e$ ($\bar\nu_e$) candidate events were observed
with an expected background of 131 (62) events for neutrino
(anti-neutrino) mode.
E776 present the observed and predicted electron energy spectra using
14 equidistant energy bins per polarity, covering the energy range
from 0~GeV to 7~GeV. In our fit we omit the first bin and combine the
second and third ones because modeling the detection efficiency at
these low energies is very difficult.  Hence, we have a total of 24
data points. We checked that we are able to reproduce well the
exclusion curve shown in fig.~4 of ref.~\cite{Borodovsky:1992pn} (if
we also use a two-flavor oscillation model), and we obtain
$\chi^2_\text{min}/\text{dof} = 31.08/22$ at the best fit point.  In
the combined analysis with other experiments we take into account
oscillations for the $\parenbar{\nu}_e$ background.

%------------------------------------------------------------------------------
\subsection{ICARUS}
\label{app:icarus}
%------------------------------------------------------------------------------

The ICARUS experiment~\cite{Rubbia:2011ft} is a neutrino beam
experiment at Gran Sasso.  The CNGS facility at CERN shoots 400 GeV
protons at a graphite or beryllium target, producing a hadronic shower
which is focused by a magnetic horn system.  The resulting neutrino
beam is mainly composed of $\nu_\mu$, having only a 2\% $\bar\nu_\mu$
contamination and a $<1$\% intrinsic $\nu_e$ component. The neutrino
spectrum ranges approximately from $0-50$ GeV, with a wide peak at
$10-30$ GeV. After traveling 732 km, they are detected in the ICARUS
T600 detector, a 760 ton liquid argon time projection chamber.
Between 2010 and 2012, the ICARUS detector observed 839 neutrino
events with energy below 30 GeV, to be compared with the expectation
of 627 $\nu_\mu$ and 3 $\nu_\tau$ charged current events, as well as
204 neutral current events. While a Monte Carlo simulation of the
experiment predicts 3.7 $\nu_e$ background events, only two were
identified~\cite{Antonello:2012pq}.  For our ICARUS simulation, we
took the $\nu_\mu$ spectrum from Ref.~\cite{Bonesini:2006ik} and
folded it with the $\nu_\mu\to\nu_e$ oscillation probability. To
assess the limit on $\nu_e$ appearance, we define the likelihood
$-2\ln(L) = 2(P-D) + 2D\log (D/P)$, where $D$ ($P$) is the total
observed (expected) number of $\nu_e$ events. Although there is a
$\sim 7$\% systematic error in the selection efficiency, we have
checked that its impact on the experiment sensitivity is negligible.

%------------------------------------------------------------------------------
\subsection{MiniBooNE}
\label{app:MB}
%------------------------------------------------------------------------------

%\subsubsection{$\nu_e$ and $\bar\nu_e$ appearance}
%-------------------------------------------------

To analyze MiniBooNE data on $\parenbar\nu_\mu \to \parenbar\nu_e$
oscillations, we use the data presented
in~\cite{AguilarArevalo:2012va}, corresponding to $6.46\times 10^{20}$
protons on target in neutrino mode and $11.27\times 10^{20}$ protons
on target in anti-neutrino mode, the same exposure as used
in~\cite{Aguilar-Arevalo:2013ara}. We follow the analysis strategy
outlined in the supporting on-line
documentation~\cite{AguilarArevalo:2012va}.  For each set of
oscillation parameters we consider, we compute the expected neutrino
and anti-neutrino energy spectra by weighting the unoscillated Monte
Carlo events from MiniBooNE's data
release~\cite{AguilarArevalo:2012va} with the oscillation
probabilities. For each simulated event, we take into account the
individual neutrino energy and the distance between the neutrino
production and detection vertices. We allow both signal and background
neutrinos to oscillate including the ``wrong-sign contamination'' of
the signal.  In computing the log-likelihood for any given parameter
point, we take into account statistical and systematic uncertainties
by using the covariance matrix published by
MiniBooNE~\cite{AguilarArevalo:2012va}. The likelihood is given by
\begin{align}
  -2\ln(L) = (\mathbf{D}-\mathbf{P})^T S^{-1} (\mathbf{D}-\mathbf{P}) \,,
  \label{eq:MB-likelihood}
\end{align}
where $\mathbf{D}$ ($\mathbf{P}$) is a column vector containing the observed
(predicted) number of events including background, and $S$ is the
covariance matrix (see Ref.~\cite{AguilarArevalo:2012va} for
details). Note that this covariance matrix includes correlations
between the $\parenbar\nu_e$ and $\parenbar\nu_\mu$ event samples  as
well as between neutrino and anti-neutrino data (if analysed
together). There are 11 bins for $e$-like data and 8 bins for
$\mu$-like data, for neutrino and anti-neutrino mode, each.

The $\pbn_\mu$-data are used to determine the normalization of the
beam flux. In principle, we should therefore include also oscillations
in the $\parenbar\nu_\mu$ disappearance sample. The impact of
$\parenbar\nu_\mu$ oscillation on the fit to the $\parenbar\nu_e$ data
would then be taken into account by the covariance matrix. This is
problematic because it would prevent us from combining, without
double-counting, our results with the independently obtained ones from
MiniBooNE's dedicated $\parenbar\nu_\mu$ disappearance searches (see
below).  Using the $\parenbar\nu_\mu$ data
from~\cite{AguilarArevalo:2012va} directly for a disappearance
analysis is also not possible because the corresponding
$\parenbar\nu_\mu$ prediction from~\cite{AguilarArevalo:2012va} has
been obtained from a Monte Carlo simulation in which some parameters
have been tuned to the data assuming no $\parenbar\nu_\mu$
disappearance. We have therefore decided to follow the MiniBooNE
collaboration and \emph{not} to include oscillations of
$\parenbar\nu_\mu$ in the appearance analysis.  We have verified
numerically that this has a negligible impact on the fit results once
external limits on $|U_{\mu 4}|$, $|U_{\mu 5}|$ are taken into
account.

A further subtlety arises for the kaon-induced backgrounds. In
MiniBooNE, these are predicted using information from the SciBooNE
detector~\cite{Cheng:2011wq}, which operated in the same beam, but at
a shorter baseline. To account for this, we rescale the kaon-induced
backgrounds in MiniBooNE by the ratio of the oscillation probabilities
in MiniBooNE and in SciBooNE. Since kaon-induced backgrounds make only
a small contribution to MiniBooNE's total error budget, the effect of
this rescaling on the fit results is, however, very small.

Due to the correlation with the muon-like events it is not straight
forward to assign a number of degrees of freedom to the appearance
search without double counting the muon data, which are used also in
the separate disappearance analysis. We have adopted the following
prescription. Eq.~\eqref{eq:MB-likelihood} can be written as
\begin{equation}\begin{split}\label{eq:MBchisq}
    \chi^2 = -2 \ln(L) & =
    d_e^T M_{ee} d_e +
    2 d_e^T M_{e\mu} d_\mu +
    d_\mu^T M_{\mu\mu} d_\mu
    \\
    &= (d_e + \delta)^T M_{ee} (d_e + \delta) + C
\end{split}\end{equation}
where $d_e$ ($d_\mu$) are the $e$ ($\mu$) components of the vector
$(\mathbf{D}-\mathbf{P})$, $M\equiv S^{-1}$, and $M_{\alpha\beta}$
denotes the corresponding sub-blocks of the matrix $M$. In
Eq.~\eqref{eq:MBchisq} we have defined
\begin{equation}\label{eq:def-C}
  \delta \equiv  M_{ee}^{-1} M_{e\mu} d_\mu
  \quad\text{and}\quad
  C \equiv d_\mu^T \left( M_{\mu\mu} -
  M_{\mu e} M_{ee}^{-1}M_{e\mu} \right) d_\mu
  = d_\mu^T (S_{\mu\mu})^{-1} d_\mu \,,
\end{equation}
where $(S_{\mu\mu})^{-1}$ is the inverse of the $\mu\mu$ sub-block of
$S$.\footnote{Note that $(S_{\mu\mu})^{-1}$ is different from
  $M_{\mu\mu}$, the latter being the $\mu\mu$ sub-block of the inverse
  of $S$.}  Hence, we have block-diagonalized the covariance
matrix. The shift $\delta$ corresponds to the impact of the $\mu$-like
data on the normalization of the $e$-like flux. The two terms in
Eq.~\eqref{eq:MBchisq} should be statistically independent and
approximately $\chi^2$ distributed. For the MiniBooNE appearance
analysis we therefore use $\chi^2_\text{MB,app} \equiv \chi^2 - C$,
and assign 22 dof to it (for combined neutrino/anti-neutrino
data). The last equality in Eq.~\eqref{eq:def-C} shows explicitly that
$C$ does not depend on oscillation parameters, since we neglect the
effect of oscillations on $d_\mu$. With this method we obtain GOF
values which are in reasonable agreement with the numbers obtained by
the collaboration: our results for $\chi^2_\text{min}/$dof (GOF) for
neutrino, anti-neutrino, combined data are 14.2/9 (11\%), 6.5/9
(69\%), 32.9/20 (3.5\%), respectively, compared to the numbers
obtained in~\cite{AguilarArevalo:2012va} 13.2/6.8 (6.1\%), 4.8/6.9
(67.5\%), 24.7/15.6 (6.7\%). Note that in~\cite{AguilarArevalo:2012va}
the number of dof and GOF have been determined by explicit Monte Carlo
study and also a different energy range has been used to obtain those
numbers.

%\subsubsection{$\nu_\mu$ and $\bar\nu_\mu$ disappearance}
%--------------------------------------

\bigskip

For our MiniBooNE $\nu_\mu$ disappearance analysis, we use the
neutrino mode data from~\cite{AguilarArevalo:2009yj}. As in appearance
mode, we compute the expected event spectra for each parameter point
by using MiniBooNE's Monte Carlo events. Since backgrounds are very
small for this analysis, we do not need to take them into account. For
each set of oscillation parameters, we choose the overall
normalization of the spectrum in such a way that the total predicted
number of events matches the number of observed events in MiniBooNE,
\textit{i.e.}, we fit only the event spectrum, not the normalization.
The log-likelihood is obtained in analogy to
Eq.~\eqref{eq:MB-likelihood} and thus takes into account systematic
uncertainties and correlations between different energy bins.

%------------------------------------------
%\subsubsection{$\bar\nu_\mu$ disappearance}
%------------------------------------------

In the analysis of $\bar\nu_\mu$ disappearance data, we follow the
combined MiniBooNE/SciBooNE analysis from~\cite{Cheng:2012yy}. As for
the appearance and $\nu_\mu$ disappearance analyses, we use public
Monte Carlo data to compute the predicted event spectra. We take into
account oscillations of both the signal and the background, and we
compute the log-likelihood again in analogy to
Eq.~\eqref{eq:MB-likelihood}.

%------------------------------------------------------------------------------
\subsection{MINOS}
\label{app:minos}
%------------------------------------------------------------------------------

Our analysis of MINOS neutral current (NC) and charged current (CC)
interaction is based on $7.2 \times 10^{20}$ protons on target of NC
data presented by the collaboration in Ref.~\cite{Vahle:2010} (see
also~\cite{Adamson:2010wi, Adamson:2011ku}) and $7.25 \times 10^{20}$
protons on target of CC data published in~\cite{Adamson:2011ig}.  All
data was recorded in neutrino mode, \textit{i.e.}, the beam consists
mostly of $\nu_\mu$ and only small contaminations of $\bar\nu_\mu$,
$\nu_e$ and $\bar\nu_e$.

We have implemented the properties of the NuMI beam and the MINOS
detector within the GLoBES framework~\cite{Huber:2004ka,
  Huber:2007ji}, using results from the full MINOS Monte Carlo
simulations as input wherever possible. In particular, we use
tabulated Monte Carlo events~\cite{Sousa:privcom} to construct the
detector response functions $R^\text{ND}(E^\text{true}, E^\text{rec})$
and $R^\text{FD}(E^\text{true}, E^\text{rec})$ for neutral current
events in the near detector (ND) and the far detector (FD),
respectively.  $R^\text{ND}$ and $R^\text{FD}$ describe the
probability for a neutrino with true energy $E^\text{true}$ to yield
an event with reconstructed energy $E^\text{reco}$. We include the CC
$\nu_\mu$ and CC $\nu_e$ backgrounds to the NC event sample (beam
intrinsic as well as oscillation induced), as well as the small NC
background to the CC $\nu_\mu$ event sample. The number of charged
current interactions is predicted using the simulated NuMI
flux~\cite{Bishai:privcom}, the cross sections calculated
in~\cite{Messier:1999kj, Paschos:2001np, Andreopoulos:2009rq}, and a
Gaussian energy resolution function with width $\sigma_E^\text{CC} /
E^\text{true} = 0.1 / \sqrt{E^\text{true} / \text{GeV}}$ for the CC
event sample and $\sigma_E^\text{CC-bg} / E^\text{true} = 0.16 + 0.07
/ \sqrt{E^\text{true} / \text{GeV}}$ for the CC background in the NC
event sample. The parameters of the energy resolution function, as
well as the efficiencies, have been tuned in order to optimally
reproduce the unoscillated event rates predicted by the MINOS Monte
Carlo.  We emulate the baseline uncertainty due to the non-vanishing
length of the MINOS decay pipe by smearing the oscillation
probabilities with an additional Gaussian of width $\tilde\sigma^2 =
(2.0~\text{GeV} - 3.0 \times E^\text{true})^2$ and setting the
effective distance between near detector and neutrino source to 700~m.
This value as well as the parameters of the smearing function have
been obtained numerically from the requirement that 2-flavor
oscillation probabilities at the near detector computed with a more
accurate treatment of the decay pipe are well reproduced for various
mass squared differences of order eV.

We compute neutrino oscillations in MINOS numerically using a full 4-
or 5-flavor code that includes all relevant mixing parameters as well
as CC and NC matter effects.  In the fit, we predict the expected
number of events at the far detector for a given set of oscillation
parameters by multiplying the observed number of near detector events
in each energy bin with the simulated ratio of far and near detector
events in that bin. Each event sample is divided into 20 energy bins
with a width of 1~GeV each, covering the energy range from 0 to
20~GeV. As systematic uncertainties, we include in the analysis of the
NC (CC) sample a 4\% (10\%) overall normalization uncertainty on the
far-to-near ratio, separate 15\% (20\%) uncertainties in the
background normalization at the far and near detectors, a 3\% (5\%)
error on the energy calibration for signal events and a 1\% (5\%)
error on the energy calibration for background events. For the NC
analysis, the systematic uncertainties are based on the information
given in~\cite{Adamson:2010wi}, for the CC analysis they have been
tuned in order to reproduce the collaboration's fit with reasonable
accuracy, while still remaining very conservative.

% -----------------------------------------------------------------------------

\bibliographystyle{JHEP}
\bibliography{refs}

\end{document}